\documentclass[11pt]{article}
\usepackage{graphicx}

\begin{document} 
\pagestyle{headings}

\begin{figure}
\end{figure}

\title{The Warped HI Layer of the Outer Galaxy}
\author{T. Voskes}
\begin{center}
\vskip 2cm
\Large {\bf The Warped HI Layer of the Outer Galaxy}
\vskip .901cm
{\bf T. Voskes}
\vskip 1.501cm 
\large {\bf M.Sc. Thesis, University of Leiden}
\vskip .501cm
%November, 1999
\end{center}

%\maketitle

\pagebreak
\vspace{1cm}
\begin{center}
{\large {\bf Abstract}}
\end{center}
%\begin{abstract}
%\vspace{1cm}
Using the Leiden/Dwingeloo Survey (Hartmann \& Burton
1997~\cite{ha:gal}) of the Galactic sky north of $\delta=-30^\circ$ as
the principal component of a composite data cube, the structure of the
warped HI layer of the outer Galaxy was displayed by converting the
data cube from heliocentric $(l,b,v)$--coordinates to galactocentric
$(R,\theta,z)$--coordinates. We masked out known high--velocity--cloud
complexes which might otherwise have contaminated the resulting
description of the warped layer. We considered analogous displays
under controlled circumstances by using as input medium a simulation
of the same complete ($l,b,v$) data cube corresponding to a modeled
HI Galaxy of known morphology and kinematics. By varying this
artificial input we were able to check our method of converting the
data cube and to put constraints on the global parameters that have
been used to explain different asymmetries in the heliocentric data.
%\end{abstract} 
 
\vspace{4cm}
{\small {This paper was written by Tom Voskes, in fulfillment of the requirements for a Master of Science degree at the University of Leiden, under the supervision of W. Butler Burton. The thesis defense took place in November, 1999.  The thesis was given a limited distribution, but was not formally published. A brief preliminary summary of this work was published as {\it ``The Large--Scale Structure of the
Outer--Galaxy HI Layer"}, T. Voskes and W.B. Burton, 1999, in {\it
``New Perspectives on the Interstellar Medium''}, A.R. Taylor, T.L.
Landecker, and G. Joncas, editors, {\it ASP Conference Series} {\bf
168}, 375.}    A PDF file containing the complete paper, including figures, is available at ftp://ftp.cv.nrao.edu/NRAO-staff/bburton/warp/warp.pdf.  Contact: W.B. Burton, bburton@nrao.edu}   
\pagebreak

\tableofcontents

\pagebreak

\section{Introduction} 

     The early Leiden and Sydney surveys of HI in the Milky Way led to
the discovery that the gas layer in the outer reaches of our Galaxy is
systematically warped away from the flat plane
$b=0^\circ$. Subsequently, it has become clear that many galactic gas
disks are globally warped; indeed, it is difficult to argue that most
galactic gas disks are not warped. The embedded perspective which
hinders some morphological investigations of our own Galaxy is in fact
quite advantageous for studies of the warp.  The first descriptions of
the Milky Way warp as given by Burke (1957)~\cite{bk:warp}, Kerr
(1957)~\cite{ke:warp}, and Oort et al. (1958)~\cite{oo:warp} have not
had to be altered in any fundamental way as investigations based on
more modern data became available.  The early work led to the
recognition that the outer--Galaxy HI layer is warped above the
equator $b=0^{\circ}$ in the first and the second longitude quadrants,
and below it in the third and fourth; the amplitude of the warp is
larger in the northern data than in the southern; the total velocity
extent in the southern outer--Galaxy layer is larger than in the
northern; and the thickness of the gas layer in both hemispheres
increases with increasing galactocentric distance.

     Although the essentials of the original findings remained
unchanged, more recent studies by Burton \& te Lintel Hekkert
(1986)~\cite{bu:warp} using principally the Leiden/Green Bank survey
of Burton (1985)~\cite{bu:HI}, and by Diplas \& Savage
(1991)~\cite{di:warp} using the Bell Labs horn antenna survey of Stark
et al. (1992)~\cite{st:HI}, have extended the information and have
yielded a more quantitative description of the warp parameters. Also,
several investigations have pointed out certain spectral
asymmetries with respect to the generally adopted approximation of kinematic
and spatial circular symmetry to which a set of different explanations
have been sought, not always in correspondence with each other (cf. Kerr (1962)~\cite{ke:ell}, Blitz \& Spergel, 1991~\cite{bl:ell}; Kuijken, 1991~\cite{ku:ell}; and Burton, 1991~\cite{bu:ell}).

The current project further extends the earlier work by using more
sensitive data with more detailed spatial as well as kinematic
coverage; it also attempts to account for possible contamination of
the results by emission from high--velocity clouds. The data was transformed into galactocentric coordinates using the IAU standards of R$_0$=8.5 kpc and $\Theta_0$=220 km/s, and a wide range of morphological parameters was further quantified. Comparing
observationally established spectral asymmetries with respect to kinematic and
spatial circular symmetry with similar asymmetries in synthetic
spectra for a range of Galactic morphological and kinematic
situations, we attempt to identify the response of various aspects of the
spectra to small changes in these models and to put constraints on the 
parameters used in them. 

\section{Data used} 

    The project utilizes the
Leiden/Dwingeloo HI survey of Hartmann \& Burton (1997)~\cite{ha:gal},
which has a greater latitude coverage than that of the Leiden/Green
Bank survey (which was restricted to $|b| \approx 20^{\circ}$), and a
greater sensitivity $(\sigma = 0.07 {\rm K})$; the new data have a
more detailed spatial coverage ($\Delta l=30',\Delta b=30'$) as well
as a more detailed kinematic resolution ($\Delta v=1$ km/s) than the
Bell Labs data and were corrected for contamination by
stray--radiation. The Leiden/Dwingeloo data covers the sky north of $\delta =
-30^\circ$. HI data of the full southern sky are not yet available,
although the latitude range $|b|\leq 10^\circ$ at $\delta < -30^\circ$
is covered by the Maryland/Parkes survey of Kerr et
al. (1986)~\cite{ke:HI}; in the region below $\delta = -30^\circ$ where
the survey of Liszt \& Burton (1983)~\cite{li:HI} was available, it was
given precedence over the Parkes data because of its higher
resolution.  The data from the two additional sources were
interpolated onto the $(l,b,v)$ grid of the Leiden/Dwingeloo survey
and scaled to brightness temperature using the parameters of the S7
standard field given by Williams (1973)~\cite{wi:s7}. In Figure 1 the resulting all--sky data cube was displayed for two cuts through galactic longitude, $l$, and velocity, $v$, at constant galactic latitudes of $b=-2^\circ$ and $b=2^\circ$.

\vspace{-.4cm}
\begin{figure} 
\vspace{-.4cm}

\caption{Representative cuts through the all-sky composite datacube. The upper panel shows an $(l,v)$ plot at $b=-2^\circ$; the lower panel, at $b=+2^\circ$. The warp at negative latitudes in the third and fourth quadrants causes the emission to extend to larger velocities when probed below the plane $b=0^\circ$, and vice versa for the warp at positive latitudes in the first and second quadrants. The peak intensity corresponds to a brightness temperature around 135 K.  The lightest greys in these displays correspond to $T_b$=4 K, much greater than the sensitivity of the LDS, $\sigma$=0.07 K. All $(l,v)$ plots in this paper use the same grey--scale, except for those latitude--averaged.}
\end{figure}

\section{Masking high-velocity-clouds}

     The distortions by high--velocity--cloud material were addressed
by masking out the principal HVC complexes.  We note that in some
directions at intermediate latitudes the total HI emissivity
contributed by anomalous--velocity gas associated with the
high--velocity and intermediate--velocity clouds can be greater than
the conventional--velocity emissivity.  Unless this material is masked
out, the velocity--to--distance analysis in section four would have attributed it to distances calculated
according to the normal rotation velocity (unless the HVC velocity
were so extreme as to be formally forbidden in terms of the input
rotation). To test what the effect of this disturbance is on the displays of the data after conversion to galactocentric $(R,\theta,z)$ coordinates, we masked out the major HVC complexes A, C, H,
and AC, as well as the complexes WA, WB, WC, and WD (see e.g. Wakker
\& van Woerden 1997)~\cite{wa:hvc}. Complex G was partially left
behind in order not to loose any well--behaved data since it has low deviation
velocities.  Then the masked data set was subtracted from the observed
data and the result was displayed in the $(R,\theta,z)$
coordinates. The differences between the results obtained without
masking out the high--velocity clouds and those obtained after masking
were minimal. The velocities of the high--velocity cloud complexes are
largely forbidden by any reasonable rotation curve. Therefore we
conclude that the properties of the Figure 3--6 type of displays are
not significantly affected by HVC contamination.

\section{Coordinate conversion}

\begin{figure} 
\begin{center}
\end{center}
\caption{Rotational situation used in the circularly symmetric velocity--to--distance conversion of the data.}
\end{figure}

     The brightness temperatures in the masked data cube of
heliocentric coordinates $(l,b,v)$ were then transformed to HI volume
densities in galactocentric cylindrical coordinates $(R,\theta,z)$
using a kinematic model of the Galaxy that is circularly symmetric and
has a flat rotation curve with $\Theta(R{\geq}$R$_{\circ}$=8.5
kpc)=220 km/s. Figure 2 shows a face--on view from the northern side
of the Galaxy of the rotational situation. An observer at the sun's
position, embedded in the equatorial plane of the rotating Galaxy,
measures a radial velocity along its line of sight towards an object
at position ($l,b,R$) of: $v(l,b,R)$=R$_{\circ}\sin l$[$\Theta(R)$/$R$
- $\Theta$(R$_{\circ}$)/R$_{\circ}$]$\cos b$. After the
velocity--to--distance transformation was carried out, volume
densities were derived in galactocentric cylinders, assuming optical
thinness.  Since at any specified velocity range the optical depth,
$\tau(v)$, and the column density, $N_{HI}(v)$ (in number of atoms at
a particular velocity in a cylinder of cross--sectional area 1
cm$^2$), are related by $N_{HI}(\Delta v)=1.823\times 10^{18}\int
T_k\tau(v)dv$ [cm$^{-2}$], with $T_k$ the spin temperature of the gas.
For the case of optical thinness,
$T_b(v)=T_k(1-\exp^{-\tau(v)})\simeq T_k\tau(v)$, so we can derive the
volume density through the integrated intensity observed: $n_{HI}(\Delta v)\propto \int T_b(v)dv/\Delta r$ [cm$^{-3}$]. If, on
the other hand, the HI in some direction contributing at a particular
velocity is opaque ($\tau(v)\gg 1)$, the observed brightness
temperature does not reveal the number of emitting atoms, but,
instead, only a lower limit. It does indicate, however, the gas
temperature, because $T_b(v)=T_k(1-\exp^{-\tau(v)})\simeq T_k$. In
this case, the profile integral yields an interesting measure:
$\int T_b(v) dv/\Delta r\simeq T_k|\Delta v/\Delta r|$. The expression
$|dv/\Delta r|$ is called the velocity--crowding parameter. It is a
measure of how fast the line of sight distance varies for a given
radial velocity increment, according to the pertaining
kinematics. Emission from regions of high velocity crowding, that is
for a low absolute value of $dv/\Delta r$, is compressed into a short
velocity interval making the spectra tend to saturate. This effect
is particularly notable in both cardinal directions ($l=0^\circ$ and
$l=180^\circ$) and also, though somewhat less, in the directions
$l=90^\circ$ respectively $l=270^\circ$. 

The volume densities
resulting from the conversion are shown in Figures 3 through 6 in
cuts through azimuth, $\theta$, (defined such that $\theta$=$l$ for a
point at infinity) and vertical height, $z$, at constant
galactocentric radius, $R$, at azimuth intervals
$\Delta\theta=1^{\circ}$, in vertical thickness steps $\Delta z$=25
pc, in cylindrical walls of thickness $\Delta R$=250 pc.

We restricted the use of data to those latitudes that are
essential in order to display the warp to any sensible extent in
radius. Thus we used data with $b>-20^\circ$ and $b<30^\circ$. This
latitude cutoff can be seen clearly in the displays as arcs through
$z$ and $\theta$ at smaller radii. One immediately also sees the limited extent of the
Maryland/Parkes survey of Kerr et al. (1986)~\cite{ke:HI} ($|b|\leq
10^\circ$) as well as that of the Liszt \& Burton
(1983)~\cite{li:HI} survey ($|b|\leq 10^\circ$, with some
indentations at the most negative latitudes). Figures 3
through 6 show that these latitude limitations put
no major restrictions on the validity of the derived warp parameters,
as the data in the northern hemisphere up to $b=30^\circ$ seems quite
sufficient to display the warp and, due to the warp's lesser amplitude
in the southern hemisphere, so does the data down to $b=-10^\circ$.

\begin{figure} 

\vfill 
\vspace{-.4cm}
\caption{Mean HI volume densities in cylindrical cuts through the composite and HVC--masked data cube at distances of $R=12$ kpc (upper) and 14 kpc (lower). The cylindrical--coordinate $(R,\theta,z)$ data set was sampled at $1^\circ$ intervals in $\theta$ and 25--pc intervals in $z$; the velocity--to--distance transformation displayed here was circularly symmetric. The darkest grey--scale was scaled to the maximum densities at each radius, to be able to show details even at large $R$, and can be compared with $n_{HI}$ values in Figures 9 and 10.}
\end{figure}

\begin{figure} 

\vfill 

%\includegraphics[width=13cm]{figure4b.ps}
%\psfig{file=/strw11/voskes/diagrams/galdata/r18.ps,width=14cm,height=9cm} 
%\vskip 17cm
\caption{Same cuts as in Figure 3, at galactocentric distances of $R=16$ kpc and 18 kpc.}
\end{figure}

\begin{figure} 

\vfill 

%\includegraphics[width=13cm]{figure5b.ps}
%\psfig{file=/strw11/voskes/diagrams/galdata/r22.ps,width=14cm,height=9cm} 
%\vskip 17cm
\caption{Same cuts as in Figure 3, at galactocentric distances of $R=20$ kpc and 22 kpc.}
\end{figure}

\begin{figure} 

\vfill 

%\includegraphics[width=13cm]{figure6b.ps}
%\psfig{file=/strw11/voskes/diagrams/galdata/r26.ps,width=14cm,height=9cm} 
%\vskip 17cm
\caption{Same cuts as in Figure 3, at galactocentric distances of $R=24$ kpc and 26 kpc.}
\end{figure}

At $\theta=0^\circ$ the emission shows a dip. This is not an
observed feature, but a consequence of the conversion's analytics that
reach a mathematical singularity at $\theta=0^\circ$. Also, although the
displays still show non--zero HI densities at radii larger than the
last displayed at $R=26$ kpc, they are not shown here because of large systematic errors. At those radii the densities have lessened to
such extent that it is difficult to distinguish between a density
decrement with increasing radius as a galactic morphological artefact
and a density decrement along velocity dispersion wings which is converted as a progressively smaller
amount of gas originating from larger radii (Figure 12). It is further evident in the
displays that the heliocentric sampling of space
reaches finer gridpoints when getting closer to the sun by noticing
that a higher resolution is attained closer towards the anticenter.

     In order to follow the transformation process under controlled
circumstances, we have simulated the complete observed $(l,b,v)$ data
cube on the same grid by solving the radiative transfer pertaining for
an ad hoc Galaxy characterized by a plane--parallel, but warped and
flaring, gas layer. This both provides an objective control on the
conversion method used and gives at the same time useful insight in
the disturbing effects on the images in $(R,\theta,z)$ coordinates of
velocity crowding, changing degrees of optical thinness, and a finite
velocity dispersion of the gas. More on this will follow in sections
5.2.1 and 5.2.2.

\subsection{Derived galactic morphology}

After conversion, the resulting set of galactocentric
representations of the Galaxy, assuming circular symmetry, can be used
to obtain a first--order impression of its morphology. For comparison
with earlier data we computed several parameters concerning this
morphology. At the same time these parameters can be
used as the most realistic input values currently available for the
simulations of section five. It must be noted, however, that an
unavoidable error is being made here inherent to the method, when we make statements on possible galactic morphologies and kinematics as
responsible for observed asymmetries on the basis of simulated models
that took as input the parameters that were computed assuming
circular symmetry. Since all parameters were azimuthally averaged to
fit circular data and, moreover, since even in a worst case scenario the
errors will typically not exceed a few to at most ten percent, we assume that this has no important effect on the conclusions that
follow.

The parameters in the Figures 8 through 12, and 14 and 15 were computed by
fitting a Gaussian to the $z$-dependent HI density distribution for
each fixed $(R,\theta)$ combination, using the GAUSSFIT routine in
IDL. We measured for each $(R,\theta)$ combination the height of the
Gaussian, its width and the location of its center. From the
$z$ location of its center we determined the parameter plotted in
Figures 8 and 9, from its height, the one in Figures 10, 11, and 12, and
from its width, the one in Figure 14 and in Figure 15. It should be
noted that the Gaussian width corresponds to the dispersion $\sigma$ and not to a
full width at half maximum (FWHM=2.36$\ast\sigma$). The
parameter in Figure 13 was computed by integrating all $z$ values for
each point in ($R,\theta$).

It would be valuable, of course, to be able to determine the total
extent of the Milky Way's HI disk in view of its theoretical
consequences for, e.g., the shape of the Galactic dark halo and the
Local Group's dynamics, to name but a few. However, this is not easily
established. Not only is there the difficulty to distinguish in the
observed spectra between the outer edge of the Galaxy and its less
outward velocity dispersed gas (of which the exact dispersion
rate at these outer parts is difficult to determine) as is described above, but
also is it likely that errors in this determination are superseded by errors in the assumption of the Galactic kinematics as
used in the conversion method. It is clearly of influence what
type of rotation curve one uses in the velocity--to--distance
transformation; currently, there is not one
uniquely determined rotation curve found yet. Also, there remains some uncertainty in the
correct value found for R$_0$. Furthermore, as will be commented
upon in section 6.1 and 7.7, some deviations from kinematic and
spatial symmetry have important consequences for the perceived Galactic
morphology after conversion. In particular, note that it appears that
the HI density at large radii is large in the fourth quadrant
with respect to the first and that of the second quadrant is large compared to
the third. As will be shown (in section 7.7) this effect is also
reproduced with slightly elliptical HI streamlines in Galactic
simulated spectra. That the data, and the parameters deduced from it,
are nevertheless displayed in ($R,\theta,z$) coordinates, using cylindrical
symmetry and the rotation curve as above, is for comparison with
earlier data (to which this has the advantages described in the
introduction) and by lack of any better established alternative yet. The
same arguments lead to the explanation of the absence of error bars on
the parameters displayed here, as it is estimated that the systematic
errors following from the conversion method are greater than the errors introduced by fitting Gaussians or than
errors in the data itself.

\begin{figure} 

\caption{Line of sight velocities for emission from increasing radius $R$ in the direction $l=90^\circ, b=0^\circ$. The solid line is according to the kinematics used in this investigation ($\Theta(R{\geq}$R$_{\circ}$=8.5 kpc)=220 km/s), the dotted line to that in Henderson et al. (1982)~\cite{he:galpar} and Diplas \& Savage (1991)~\cite{di:warp} ($\Theta(R{\geq}$R$_{\circ}$=10.0 kpc)=250 km/s), and the dashed line corresponds to the kinematics specified by R$_\circ$=8.5 kpc, $\Theta_\circ$=220 km/s, and $\Theta(R)=\Theta_\circ(R/$R$_\circ)^{0.0382}$ as is used by Wouterloot et al. (1990)~\cite{wo:galpar}}
\end{figure}

\begin{figure} 

\vfill 

\vfill 

\caption{Vertical distance from the plane $b=0^\circ$ of the gauss-fitted peak HI density at galactocentric distances of respectively $R=12$ kpc, 16 kpc, and 20 kpc. The distribution was smoothed over $10^\circ$ in $\theta$. The central bump near $\theta =0^\circ$ is not physically significant, due to the smearing of the data in that direction (see the comments on this in the caption of Figure 14).}
\end{figure}

\begin{figure} 

\vfill 

\vfill 

\caption{Vertical distance from the plane $b=0^\circ$ of the gauss-fitted peak HI density at galactocentric distances of respectively $R=24$ kpc, 26 kpc, and 28 kpc. The distribution was smoothed over $10^\circ$ in $\theta$.}
\end{figure}

\begin{figure} 

\vfill 

\vfill 

\caption{Azimuthal distribution of HI midplane densities, measured as the gauss-fitted peak of the vertical HI density distribution, at the respective radii of $R=10$ kpc, 12 kpc, and 14 kpc. The distribution was smoothed over $10^\circ$ in $\theta$. The plots show an apparent underabundance of HI gas in those directions ($\theta\sim 0^\circ, \pm 180^\circ$) where due to velocity crowding the total amount of emitting gas is expected to be higher than what is observed by received emission. Note, specifically, the low values around $l=0^\circ$.}
\end{figure}

\begin{figure} 

\vfill 

\vfill 

\caption{Same azimuthal distribution of HI midplane densities as in Figure 10, at the radii of $R=16$ kpc, 18 kpc, and 20 kpc.}
\end{figure}

\begin{figure} 

\caption{Azimuthally averaged HI midplane densities with increasing radius for the range $20^\circ\leq\theta\leq160^\circ$ (dotted line) and for the range $200^\circ\leq\theta\leq340^\circ$ (dashed line). It is difficult, at very large radii, to distinguish between a density decrement with increasing radius as a galactic morphological artefact and a density decrement along velocity dispersion wings which is by the conversion method naturally taken as a progressively smaller amount of gas originating from larger radii.}
\end{figure}

\begin{figure} 

\vfill 

\caption{Azimuthally averaged HI surface densities ($\sigma$) in solar masses per square parsec (upper panel), for the range $20^\circ\leq\theta\leq160^\circ$ (dotted line) and for the range $200^\circ\leq\theta\leq340^\circ$ (dashed line). The lower panel gives the logarithmic version from which we deduce that the mean radial scalelength for the HI surface density in the $\theta$--averaged range $20^\circ\leq\theta\leq160^\circ$ at $R\geq 9$ kpc is 4.3 kpc and for the range $200^\circ\leq\theta\leq340^\circ$ 4.9 kpc. However if we only measure at $15\leq R\leq 25$ kpc, then we arrive at 3.7 kpc and 4.2 kpc, respectively. In view of the low densities at $R>25$ kpc, it is difficult to determine the significance of the values of $\sigma$ at those radii.}
\end{figure}

\begin{figure} 

\vfill 
\vfill 
\caption{Gaussian dispersion ($h$) of the vertical HI density distribution at distances of $R=12$ kpc, 16 kpc, and 20 kpc. The distributions were smoothed over $10^\circ$ in $\theta$. Note the increase of average scale height with increasing radius. The large values of $h$ near $\theta =0^\circ$ are spurious, being due to the elongated patch of emission in the Figures 3--6. This is caused by local gas that reaches velocities, due to the finite velocity dispersion, that correspond kinematically with radii in the far outer Galaxy. Note, that it is slightly off--center, as is the elongated patch.}
\end{figure}

\begin{figure} 

\caption{Flaring of the azimuthally averaged gaussian width ($h$) of the vertical HI density distribution for increasing galactocentric radius. The dotted line is the $\theta$--averaged value for the range $20^\circ\leq\theta\leq160^\circ$, the dashed line is the $\theta$--averaged value for the range $200^\circ\leq\theta\leq340^\circ$. Finally, the solid line is their average. The $40^\circ$--bands in both cardinal directions were omitted, because of their unrealistic high values of the scale height as explained in the caption of Figure 14. As mentioned in Figure 13, also here it is difficult to determine the significance of the values of $h$ at $R>25$ kpc, in view of the very low densities at those radii in Figure 12.}
\end{figure}

It is instructive to compare these new data with the findings of past
investigations. For instance, although the shapes of the surface
density ($\sigma$) curves of Henderson et al. (1982)~\cite{he:galpar}
in their Figure 4 and that of our Figure 13 are quite similar it can
be seen that the peak of their northern surface density profile occurs
at a radius of $\sim$14 kpc, while ours occurs at $\sim$12.5 kpc. This
difference is caused by the choice of the assumed Galactic kinematics
in the velocity--to--distance transformation as is illustrated in
Figure 7. This scale effect, which is more pronounced for radii nearby to the
solar radius, is squared when it enters their analysis of the surface
density, using $\Theta(R{\geq}$R$_{\circ}$=10.0 kpc)=250 km/s, and
leads to an underestimate relative to the analysis we made, using
$\Theta(R{\geq}$R$_{\circ}$=8.5 kpc)=220 km/s. Thus the difference in the
magnitude of the density, with the current findings coming to a more
than 2 solar masses per square parsec higher value than those of
Henderson et al. for radii up to about 17 kpc, is at least in part
accounted for. This is clearly also the case in the logarithmic plot
of $\sigma$ in Figure 8 of Diplas \& Savage (1991)~\cite{di:warp},
who used the same kinematics as Henderson et al. Also, they show in
their table 1 that derived scale lenghts are expected to be greater
when using the kinematics we used than using theirs. They point out
that for the Henderson et al. data they estimate $r_s$=2.08 kpc (for
$\Theta(R{\geq}$R$_{\circ}$=10.0 kpc)=250 km/s), while they arrive at
$r_s$=2.49, 2.60, and 3.98 kpc for values of $\theta$ of $50^\circ,
90^\circ$ respectively $130^\circ$ corresponding to $r_s$=3.14, 2.74, and
5.55 kpc for $\Theta(R{\geq}$R$_{\circ}$=8.5 kpc)=220 km/s. These three specific values can not be compared directly to the
$\theta$--averaged values we found as explained in the caption of
Figure 13, though it is clear that ours are significantly higher than
the one found from the Henderson et al. data. Since the shape, and
particularly the height, of the Galactic surface density curve is very
sensitive to the choice of range for averaging in $\theta$, as can
clearly been seen in Figures 10 and 11. In the case of the work done
by Henderson et al. this range was $30^\circ\leq\theta\leq150^\circ$
(northern) respectively $210^\circ\leq\theta\leq330^\circ$ (southern) where
we used $20^\circ\leq\theta\leq160^\circ$
respectively $200^\circ\leq\theta\leq340^\circ$, some of the difference in
surface densities along Galactic radius may also be accounted for by
this effect, although the lower densities toward the cardinal
directions might in fact argue for a bias in the Henderson et al. data
towards higher values with respect to ours, instead of lower. That this effect
is significant and can be attributed to velocity crowding alone is
shown in the lower panel of Figure 22 in section 5.2. This is also the
direct reason for us to leave out a band of $40^\circ$ in $\theta$
in both center and anticenter direction in our analysis. The data as
analyzed by Wouterloot et al. (1990)~\cite{wo:galpar} is more directly
comparable to the current results, as can be found in their
Figure 5b, due to the small difference in kinematic distance
determination between their method, using R$_\circ$=8.5 kpc,
$\Theta_\circ$=220 km/s, and
$\Theta(R)=\Theta_\circ(R/$R$_\circ)^{0.0382}$, and ours, as can again
be seen in Figure 7. They find for the ranges
$120^\circ\leq\theta\leq170^\circ$ (northern)
respectively $190^\circ\leq\theta\leq240^\circ$ (southern) equivalent, though
slightly lower, magnitudes than ours, where due to a bigger overlap with
the region that is most affected by underestimation of density through
velocity crowding one would expect these magnitudes to be lower
limits, rather than upper. This is almost certainly the case for the
Galactic surface density curve shown in their Figure 5e, where they
averaged over the complete second and third quadrants and thus
included the minimum around $l=180^\circ$. As therefore expected, it
displays lower $\sigma$, especially at radii nearer to R$_0$=8.5 kpc,
which can be understood when taking into account the progressively larger relative
minimum in the Figures 10 and 11 for values of $R$ nearer to the solar
radius. Their value of radial scalelength, $r_s=4.0$ kpc, is, again,
equivalent to ours, though slightly lower, in spite of the suppressed
density magnitudes at radii close to R$_0$, which would make it
larger. This could be attributable to the fact that at larger radii
the difference between the kinematics becomes more pronounced (Figure
7), causing their $\sigma$ to decline faster.

The flaring of the thickness of the HI layer is demonstrated in Figure
15. Also here it can be seen, from Figure 14, that the exact shape
and, in particular, the height of the flaring curve is sensitive to
the applied range of $\theta$--averaging. It is remarkable how the
data from Wouterloot et al., in their Figure 9, leads to a much
smaller thickness near $R=10$ kpc of a half width to half maximum of
about 200 pc, which roughly equals 170 pc (FWHM=2.36$\ast\sigma$) in
Gaussian width ($\sigma$), compared to our $\sim$450 pc in
$\sigma$. However, Diplas \& Savage find, using the data of Henderson
et al., for a $\theta$--averaging over
$90^\circ\leq\theta\leq130^\circ$ about $\sigma$=500 pc and for
$210^\circ\leq\theta\leq250^\circ$ about $\sigma$=400 pc, both at
$R=12$ kpc, in good agreement with the data here. Their plots of the
flaring of the thickness for different particular $\theta$ values in
this figure, with their wide range in thickness at any particular
radius, illustrate the impact of the range of $\theta$ included on
the flaring curve, as was mentioned above. We choose not to include a
band of $40^\circ$ in $\theta$ in both center and anticenter direction
in our analysis, since we can see that in these regions the values of
$h$ are unrealistically high due local velocity dispersed gas, explained in
the caption of Figure 14, as is demonstrated in Figure 22, upper panel,  in section
5.2. For all the plots showing the flaring of the thickness, the
flaring continues to rise towards higher radii. Only our new data
shows a slowing of rise in flaring at very large radii ($R>23$ kpc),
although one should be careful about the interpretation of the
analysis at these large distances from the Galactic center in view of
what was said before about systematic errors.

When looking at the vertical distances from the plane $b=0^\circ$ of
the gauss-fitted peak HI density at very large galactocentric
distances in Figure 9 the attention is drawn to a feature, which is
referred to in the literature as scalloping (e.g. Kulkarni et al. (1982)~\cite{ku:scal}), that is most pronounced
in the 3$^{rd}$ and 4$^{th}$ quadrants. As mentioned before,
carefulness is required for the interpretation of the
reality--content of this feature at such large radii. Closer
inspection reveals that about 10 periods can be distinguished for a
full $360^\circ$ in azimuth, the same number that was found by
Kulkarni et al. The scalloping is further
remarkable, because it sustains the same regularity through several
kpc in galactocentric radius. Currently, it is uncertain,
what its cause is. In
any case this effect is additional to the global parameters we
consider here and shall not be pursued by us any further.

\subsection{Computed input parameters}

We have constructed a complete set of $(l,b,v)$ spectra based upon an
HI brightness temperature distribution according to known input
galactic morphology and kinematics. More on this will follow in
section 5. Both for the purpose of having a realistic controlled input
medium for the heliocentric--to--galactocentric coordinates conversion
and for the purpose of showing observational consequences of various
different galactic models that have been put forward as a possible
explanation of observed spectral asymmetries, it is valuable to have
close--to--reality input parameters. Some of these are taken from
the literature, others have been computed by using this project's data and
the parameters calculated in the subsection above. We will mention
those of the last type here and explain the method by which they were
calculated.

\begin{figure} 

\caption{Solid lines trace the sine-fitted warp heights (see text) for the first and second quadrant together (upper) respectively third and fourth (lower). The dashed lines are the mathematical fits that are used in the model.}
\end{figure}

The radial scale length was taken as an average of the four values that were found from 
the lower panel in Figure 13. It was derived to be 4.3 kpc. The vertical scale height in the
solar neighborhood was deduced from Figure 15. This
is a Gaussian scale factor and it was computed to be 0.40 kpc. The
rate of flaring was also taken, azimuthially averaged, from the same
plot and amounted to 25 pc in Gaussian width of vertical thickness for
every kpc radially outward beyond $R=$9 kpc, the radial edge of the assumed
flat and of constant density disk. This central disk HI density was
measured at a radius of $R=9$ kpc and averaged over all
$\theta$ values except for $20^\circ$ regions centered on both the
anticenter and center directions in order to avoid an underestimation
due to velocity crowding. We found 0.26 cm$^{-3}$, which is a lower limit. At a certain point the simulation makes use
of a sinusoidal warp that is linearly rising beyond $R=9$ kpc. That
this is a fair first approximation can be seen in Figure 8. To
determine the rate of rising of the warp we calculated the magnitude
of a sine curve that maps out the same surface in a plot either above or
below the plane $b=0^\circ$. These thus determined sine--fitted
warp heights for increasing radius can be found as the solid lines of
Figure 16. Their increasing rate was averaged to give 110 pc per kpc
radially outward from $R=9$ kpc. A better approximation involves an
asymmetric warp that distinguishes a part at latitudes north of the
plane $b=0^\circ$ and one south of it. This approximation increases
not linearly but according to the mathematical fits shown with dashed
lines in Figure 16.

In the original $(l,b,v)$ data the warp maximum and warp minimum
were defined as the longitudes at which the highest respectively lowest
latitudes were attained. This was measured for a velocity--integrated
$(l,b)$ plot at $-220 < v < -100$ km/s for the warp at positive latitudes
in the first and second quadrant to include only gas from the outer
Galaxy, as well as for $100 < v < 220$ km/s for the warp at negative
latitudes in the remaining two quadrants. This resulted in
$l=103^\circ$ for the warp maximum and $l=262^\circ$ for the
warp minimum. Because of the smaller than $180^\circ$ spacing between
the maximum and the minimum and the known folding--back property of
the warp at the southern hemisphere, which is prominent in the
velocity range that was sampled here, the measurement was repeated for
the range $60 < v < 100$ km/s, which is representative for distances more
towards those where the southern warp reaches its minimum, shown in
Figure 16. This gave for the warp at $b<0^\circ$ a quite shallow
minimum that extended from $l\sim 240^\circ$ to $l\sim 290^\circ$. A
warp maximum for $l=100^\circ$ implies for the models a galactocentric
warp maximum at $\theta\simeq120^\circ$ spaced from the warp minimum
by $180^\circ$, assuming that this maximum is observed at $R\sim20$
kpc or more. However, judging by the line of nodes in Figures 8 and 9
one would come to a warp maximum at $\theta\simeq95^\circ$. As, from
the same figures, it is clear that both orientations represent only in part
the real situation, since the shape of the warp is not perfectly
sinusoidal, both will be used in the simulations of section five and
seven.

\section{Modeling under controlled circumstances}

We have constructed a complete set of $(l,b,v)$ spectra based upon an
HI brightness temperature distribution according to known input
Galactic morphology and kinematics. This simulation provides a
strong check on the analysis procedures, when taken with identical
kinematics as those specified in the method used for converting the
data to ($R,\theta,z$) coordinates. Changing the input models, it is
furthermore possible to monitor the influences this has on both the
appearance of the spectra, and on the appearance of plots in
($R,\theta,z$), that were made, while still using the conversion we
use in the real data's case. To be able to include demonstrations of
the effects of any possible kinematic ellipticity or other forms of
asymmetry, we have used as an input model the general case of a
Galaxy with kinematics described by motions in closed elliptical
streamlines. That it is reasonable to do so, even though in a
non--spherical potential there are no dynamical stable and closed
orbits, can be thought of when one realizes that a collection of
HI clouds on independent and randomly chosen non--stable orbits are
likely to collide and thus to cause shocks, for which there is no
observational evidence. The kinematics were therefore chosen without a
formal dynamical foundation, with an ellipticity, $\epsilon$, defined
for the rotational velocity, $\Theta$, of any object at a certain
position angle $\psi$ (which is $\alpha+180^\circ$ in the case of the
sun; see Figure 17). This $\Theta$ is decomposed into a velocity along
the x-axis (defined by the ellipse major axis), $\Theta_x$, and a velocity
along the y-axis, $\Theta_y$, where
$\Theta_x=-\Theta_0(1+\epsilon/2)\sin\psi$ and
$\Theta_y=\Theta_0(1-\epsilon/2)\cos\psi$.

\begin{figure} 
\caption{General elliptical rotational velocities used in the simulation models, $\Theta_x$ has a negative value, $\Theta_y$ a positive one. Note that in the case $\epsilon=0$ and $\alpha=0^\circ$, it will follow that $\psi=\theta$.}
\end{figure}

\subsection{Circular symmetry}

\subsubsection{Description of the model}

The simplest case we consider is that of a plane--parallel circularly
symmetric Galaxy, with differential rotation according to
$\Theta(R)=\Theta_0$[$1.0074(R/$R$_0)^{0.0382} + 0.00698$]-2.0 (km/s),
for $R<$R$_0$, and $\Theta(R)=\Theta_0$, for $R\geq$R$_0$. The gas
layer shows flaring, but is not warped. The midplane is specified by
the plane $b=0^\circ$ and has a volume density of 0.30 cm$^{-3}$, a
little higher than what was found in section 4.2 to compensate for the
underestimation of this parameter due to the fact that not everywhere
in the Galaxy along the line of sight the opacity is truly far smaller
than 1. That this expectation is justified can be seen in the middle
panel plot of Figure 21, which bears a fair resemblance to that of
Figure 12. The volume density is constant at the midplane up to $R=9$
kpc, but at larger $R$ it declines with a scale length of $r_s=$4.3
kpc. Moreover, it also declines with vertical $|z|$ height, with a
scale height of $h_s$=0.40 kpc. Due to the flaring of this thickness,
this value of $h_s$ increases with 0.025 kpc for every kpc in radius
beyond $R=9$ kpc. All these parameters were derived using the real data
after a circularly symmetric conversion as described in section 4.2. The
model Galaxy is truncated at $R=30$ kpc. The gas was defined to have
a velocity dispersion of $\sigma=7.5$ km/s and a spin temperature of
$T_k=135$ K.

\subsubsection{Simulation results}

The complete observed $(l,b,v)$ data cube is then simulated on the
same grid by solving the radiative transfer pertaining for the ad hoc
Galaxy, as described above. Figure 18 shows the resulting set of
spectra at $b=0^\circ$ in a display through $l$ and $v$. Note that,
apart from the scale factors mentioned above, the input model did not
have any structure and that thus all artefacts in this ($l,v$) plot
are due to effects of velocity crowding, i.e. to changing degrees of optical
thinness, and a finite velocity dispersion of the gas. Furthermore, the line of sight kinematic point of symmetry with
respect to the outer Galaxy is entirely coincident with the location
of the sun. Thus, the most extreme velocities occur at galactic
longitudes of $l=90^\circ$ respectively $l=270^\circ$ and show a
drop--off in magnitude that is symmetrical with respect to the
velocity minima at the center ($l=0^\circ$) and anticenter ($l=180^\circ$)
directions. Note also, how in the 2$^{nd}$ quadrant some emission is
present at formally forbidden positive velocities, as is some emission
in the 3$^{rd}$ at negative velocities, due to the velocity dispersion
of nearby gas.

\begin{figure} 

\caption{Simulated analogue at $b=0^\circ$ of Figure 23, upper panel, for a model with complete circular symmetry as described in the text. As in Figure 1, the peak intensity of the display corresponds to a brightness temperature of 135 K; the lightest greys, to $T_b$=4 K.}
\end{figure}

\subsection{Circular kinematics with warp}

Since the presence of a warp in the Galactic HI layer will
alter the appearance of the velocity cutoff at $b=0^\circ$ in those
directions were the line of sight will leave the gas layer where it
warps away maximally from the plane at radii smaller than those for
which it reaches its extreme velocities, it is instructive to compare
the observed ($l,v$) plot at $b=0^\circ$, with an ($l,v$) plot of a
model Galaxy described as in subsection 5.1.1, only now with a linear
sinusoidal warp. The shape of the warp is such that a line through the
midplane along a cylindrical wall through $\theta$ and $z$ at constant
radius $R$ is a perfect sine curve. The magnitude of the maximum of the
warp starts at zero at $R=9$ kpc and increases from there with 110 pc
for every kpc outwards, as was derived from the real data in section
4.2. For the simulations in this section we put the warp maximum at
$\theta=95^\circ$.

\subsubsection{Conversion controlling}

\begin{figure}

\vfill 
%\includegraphics[bb=70 370 550 720, width=13cm, height=9cm]{figure19b.ps}
%\psfig{file=/strw11/voskes/diagrams/circwarp/lvbintabs10m95.ps,width=14cm,height=9cm} 
%\vskip 17cm
\caption{(l,v) plot at $b=0^\circ$, for the simulated data with circular kinematic symmetry, but a symmetrical warp with its maximum at $\theta=95^\circ$ (upper panel), as well as a latitude integrated version for $|b|\leq 10^\circ$ (lower panel), to show the effects this has on the influence of the warp. As in Figure 1, the peak intensity of the upper--panel display corresponds to a brightness temperature of 135 K and the lightest grey--scale corresponds to $T_b$=4 K. The lower--panel display was averaged over its $b$ values and has a deviating grey--scale scaled between minimum and maximum intensities.}
\end{figure}

In Figure 19 an ($l,v$) plot of the sort of Figure 18 was displayed,
this time including a linearly rising sinusoidal warp. For comparison
with a plot shown in Figure 23 in section 6.1 a latitude integrated
version for $|b|\leq 10^\circ$ was displayed as well. The upper panel shows indentations at the outer--Galaxy
velocities in the heliocentric $l$ directions where we expect the
maximum warp along galactocentric $\theta$. This complete $T_b(l,b,v)$
data cube was processed in the same manner as the observed data cube to
yield volume densities in cylindrical cuts through galactocentric
coordinates of the sort displayed in the Figures 3--6. Since,
in this case the input kinematics are, for certain, conveniently
identical to those used in the velocity--to--distance processing, this
should render a Galactic morphology that is consistent with the
input model and, furthermore, draw our attention to any vagaries that
can only have been inflicted by the method of converting the data
cube and that can thus also be expected to blur our view of the
Galaxy, displayed in the Figures 3--6. The result of this conversion
is displayed in galactocentric cylindrical cuts in Figure 20 for the
constant radii $R=16$ kpc and $R=26$ kpc.

\begin{figure} 
\vfill 
%\includegraphics[width=13cm, height=8cm]{figure20b.ps}
%\psfig{file=/strw11/voskes/diagrams/circwarp/r26.ps,width=14cm,height=9cm} 
%\vskip 17cm
\caption{Mean HI volume densities in cylindrical cuts through the simulated 
data cube with circular kinematic symmetry, but a symmetrical warp with its maximum at $\theta=95^\circ$ at galactocentric distances of $R=16$ kpc (upper) and 26 kpc (lower).
The cylindrical--coordinate $(R,\theta,z)$ data set was sampled at $1^\circ$ 
intervals in $\theta$ and 25--pc intervals in $z$; the velocity--to--distance 
transformation displayed here was circularly symmetric. The darkest grey--scale was scaled to the maximum densities at each of both radii, to be able to show details even at $R$=26 kpc. Compare, for the exact derived densities, Figure 21 (middle panel). For reference a line tracing the galactic plane at $b=0^\circ$ was drawn.}
\end{figure}

\begin{figure} 

\vfill 

\vfill 

\caption{Vertical distance from the plane $b=0^\circ$ of the Gauss-fitted peak HI density of the simulated 
data cube with circular kinematic symmetry, but a symmetrical warp with its maximum at $\theta=95^\circ$ at a distance of $R=16$ kpc (upper), to be compared with Figure 8, a $\theta$-averaged ($20^\circ<\theta<140^\circ$) HI midplane density with increasing radius (middle) (comparable with Figure 12) and the flaring of the azimuthally averaged Gaussian width of the vertical HI density distribution (lower), comparable with Figure 15. The parameters were calculated in the same manner as their real--data counterparts, that is, after conversion of the data to galactocentric coordinates.}
\end{figure}

\begin{figure} 

\vfill 

\caption{Gaussian width ($h$) of the vertical HI density distribution (upper) of the simulated 
data cube with circular kinematic symmetry, but a symmetrical warp with its maximum at $\theta=95^\circ$ at a galactocentric distance of $R=12$ kpc (comparable with Figure 14) and azimuthal distribution of HI midplane densities, at $R=10$ kpc (lower) (comparable with Figure 10). Note that these parameters were calculated in the same manner as their real--data counterparts, that is, after conversion of the data to galactocentric coordinates.}
\end{figure}

\subsubsection{Resulting vagaries}

A set of morphological parameters similar to those in section 4.1
were computed using the same procedures. They are displayed in Figures
21 and 22. Several remarks can be made about the appearance of the
cylindrical cuts in Figure 20 and on these parameters. First, it can
be seen that the orientation of the warp is correct and also, at
closer inspection, that the line of nodes is exactly at
$\theta=5^\circ$ and $\theta=185^\circ$ (see also the upper panel of
Figure 21). In fact, it was tested how the conversion analytics would
deal with a warp maximum at exactly $\theta=0^\circ$, in view of the
apparent coincidence of having the sun--center line so close to the
observed line of nodes. It behaved  properly and retrieved
the exact input situation. Secondly, the gas layer has generally a
constant thickness along azimuth and shows flaring from the $R=16$ to
the $R=26$ plot, as required by the input. A non--input feature is clearly the density indentations at the center and
anticenter directions, due to a mathematical singularity in the
conversion analytics at $\theta=0^\circ, 180^\circ$. Also,
the striking patches, stretched greatly in $z$, are caused by the
processing, rather than inherent to the model. Its origin is of local
gas that reaches velocities, due to its finite velocity dispersion,
that correspond kinematically with radii in the far outer Galaxy. The
densities are then derived from the received intensities according to
the multiplefold distances. It occurs only at the cardinal directions,
as it is there where the line of sight velocities remain quite small,
even at large radii. Note that in the case of kinematic symmetry, as
is considered here, the patch is also symmetrically oriented around
both directions, whereas in the Figures 3--6 it was displaced. Compare,
in this context, for instance, the upper panel of Figure 22 with that of Figure 14. That the patch reaches greater vertical heights above
the plane in the center direction can be easily understood in terms of
the large kinematic distance that local gas emission at any particular
latitude corresponds to, with its corresponding large
$z$ height, since it needs to traverse the sun--center radius twice
first. Again, in the center direction, and in that of the anticenter
to a lesser extent as well, a vagary is observed, at $R=16$ kpc,
displaying a progressive lower volume density when approaching
$\theta=0^\circ$ (and $\theta=180^\circ$). This effect is even more
clear in the lower panel of Figure 22 and is attributable to a
non--negligible optical depth due to velocity crowding. This
concentrates all emission in a small range of frequencies and thus
causes the spectra to saturate, resulting in an underestimation of the
true volume density in spatial distribution. Finally, this
under abundance near the cardinal directions changes into an
overabundance for plots at very large radii; see the cylindrical cut
at $R=26$ kpc. By the drop in density according to the input scale
length, $r_s$, at these distances its magnitude remains only a
fraction of the inner constant density disk value, as is illustrated
in the middle panel plot of Figure 21. The contribution of emission
far into the velocity dispersion wings from gas nearer to the Galactic
center becomes non--trivial at such low
densities. Generally, the effects induced by velocity crowding and
velocity dispersion are stronger in the center direction than in
anticenter direction, because there the inner solar radius gas
emission confers at low positive and low negative velocities with the
outer--Galaxy emission through its own finite velocity
dispersion. From the middle panel of Figure 21, it can be seen
that the input density of $n_{HI}=0.3$ cm$^{-3}$ yields realistic
volume density values after conversion. The lower panel in that
figure was shown for the remarkably steep rising scale--height for
smaller radii, something that is also observed in the analysis of the
real data in this project, as well as in that of others. It is,
obviously, not according to the original input, but due to a strong
broadening of the elongated patches mentioned before when approaching
R$_0$, thus biasing the azimuthally averaged scale height to greater
values.

\section{Deviations from circular symmetry}

     There is substantial long--established evidence for bilateral
asymmetry in the outer--Galaxy gas layer. The kinematic cutoff at
$b=0^\circ$ corresponding to the far outskirts of the Galaxy occurs at
velocities some 25 km/s more extreme in the southern--hemisphere data
than in the northern (Burton, 1973~\cite{bu:lop}). Additionally, the
tightest constraint of radial velocities occurs $6^\circ$ removed from
the cardinal direction $l=180^\circ$; for $l=90^\circ$, circular
symmetry would formally forbid any positive velocities (except those
due to velocity dispersion), while, in fact, the emission peaks at
$v=+6$ km/s (Burton, 1972~\cite{bu:pro}). Early on, Kerr
(1962)~\cite{ke:ell} suggested that the apparent difference between
the inner--Galaxy northern and southern rotation curves derived from
terminal velocity data could be due to an outward motion of the
adopted local standard of rest of about 7 km/s. Blitz \& Spergel
(1991)~\cite{bl:ell} reproduced the shape of the difference between
velocity contours in the $0^\circ<l<180^\circ$ and
$180^\circ<l<360^\circ$ regions corresponding to a 1 K isotherm by
adopting a 14 km/s radial outward LSR motion and, assuming this to be
kinematically equivalent to a more naturally explained radial outward
component at the sun's location along global elliptical streamlines,
they put forward a model of gas streamlines with decreasing
ellipticity outwards, the outermost gas moving in almost circular
orbits, with an angle $\alpha$ between the sun--center line and the
semi--major axis of the gas distribution of
$\approx45^\circ\pm20^\circ$, and an ellipticity
$\epsilon\approx0.02$, which they found to be consistent with some
other observables they examined (e.g. stellar kinematics, molecular
gas). Kuijken \& Tremaine (1991)~\cite{kt:ell} and Kuijken
(1992)~\cite{ku:ell} argued that from their examination of the
kinematics of the solar neighborhood they found little or no evidence
for large--scale nonaxisymmetry. They note, however, that it would be
very difficult to detect any ellipticity, if the sun happens to reside
near a symmetry axis. These and other investigations were reviewed by
Burton (1991)~\cite{bu:ell}.

The dynamical situation responsible for the observed asymmetries
remains unclear, as does the extent to which the asymmetries might be
spatial rather than kinematic. However, it is clear that even small
deviations from kinematic symmetry are likely to have an observable
impact on the HI profiles, since their shape is very sensitive to
small variations in the streaming motions (Burton,
1972)~\cite{bu:pro}. We will consider in what follows consequences of
kinematical asymmetry only, although we note that it seems unlikely
that this would be the exclusive cause of the observational situation
in view of the common occurrence of true lopsidedness in other nearby
spiral galaxies (Baldwin et al., 1980~\cite{ba:lop}; Richter \&
Sancisi, 1994~\cite{ri:lop}). In any case, it is important to realize
that many kinematic asymmetries are necessarily joined by a spatial
counterpart (e.g. non--zero ellipticity, with the sun--center line
anywhere in between the semi--major and semi--minor axes of the
ellipsoid) and that a true structural lopsidedness, due to an $m=1$
harmonic term in the Galactic potential, is always revealed by a
perturbation in the line of sight velocity field (Swaters et al.,
1999~\cite{sw:lop}).

We do not go into a full detailed treatment in order to pinpoint a
definite Galactic morphology and its kinematics that can account for
the observed deviations from circular symmetry, but instead we want to
make use of our model simulations as a controlled input, showing the
observational implications of assumed real galactic asymmetries that
have been suggested as explanations of the observed asymmetries on
several aspects of observed spectra at the same time. In fact, this
gives us a powerful instrument for putting constraints on various
solutions to the asymmetry problem in the literature.

In order to do so, we will first distinguish a set of observed
asymmetries and establish a method to measure these asymmetries
quantitatively. Then, after having found the real data's values of
these parameters, we will adopt a list of possible morphologies and
kinematics and measure the same set of parameters in the data as
simulated according to these models. The measurements take place in an
$(l,v)$--displayed set of spectra at $b=0^\circ$. Some measurements
shall be taken as well from a $b$--integrated $(l,v)$ display, for
$|b|\leq 10^\circ$, to remove most of the warp's influence on these
displays as is demonstrated in the next section.

\begin{figure} 

\vfill 

%\includegraphics[width=13cm]{figure23b.ps}
%\psfig{file=/strw11/voskes/diagrams/lbvdata/lvbintabs10.ps,width=14cm,height=9cm} 
%\vskip 17cm
\caption{(l,v) plot at $b=0^\circ$ (upper panel) as well as a latitude integrated version for $|b|\leq 10^\circ$ (lower panel), which can be considered free of any influence by deviations from the galactic plane. Like in Figure 1, the peak intensity of the upper panel corresponds to a brightness temperature around 135 K and the lightest grey--scale corresponds to $T_b$=4 K. The lower panel was averaged over its $b$ values and has a deviating grey--scale scaled between minimum and maximum intensities.}
\end{figure}

\subsection{Asymmetries in the data}

We define spectral asymmetry as any feature or property of the
$T_b$ spectral $(l,v)$ images, insofar as not corresponding to those
according to simple planar circularly symmetric rotation, with the
assumed flat rotation curve as described in section 4.  

In the $(l,b,v)$ data we choose to monitor the following set of parameters, when we reproduce galactic spectra according to different
models. Most of these have been used before in investigations on
Galactic asymmetry and some have been added by us, not always as an
asymmetry, but also to put restrictions on the various explanations of
observed asymmetry. We explain how we proceed to measure these
phenomena and give the according values for the real data.

\begin{enumerate}
\item The kinematic or spatial cutoff appears to occur at more extreme velocities in the part of the hemisphere accessible from the south than that from the north.
\item The maximum velocities attained in the north and the south are not centered on $l=90^\circ, 270^\circ$ respectively, as would be expected in the symmetrical case.
\item The shapes of the extreme velocity curves in the north and the south differ.
\item The tightest velocity constraint towards the anticenter is displaced by a few km/s into the third quadrant.
\item The centroid or middle of this velocity distribution still appears centered on $l=180^\circ$.
\item The tightest velocity constraint towards the center is not displaced as it is towards the anticenter.
\item The centroid of the velocity distribution around $l=0^\circ$ is significantly displaced to positive velocities.
\item Emission extends to higher positive velocities in the second quadrant than to corresponding negative velocities in the third.
\item Finally, it appears that the velocity integrated column depth is large in the fourth quadrant with respect to the first, and that of the second quadrant is large compared to the third.
\end{enumerate}

We can further make remarks on the apparent asymmetries, as observed in the real data after conversion under the assumption of circular symmetry, in the ($R,\theta,z$) frame, as can be seen in the Figures 3 through 6. They were not measured quantitatively.

\begin{enumerate}
\item In general the South is observed to be endowed with a progressively higher abundance of gas with increasing radii than the North.
\item An elongated patch of gas in the center direction, reaching high $z$, is observed to be displaced from $l=0^\circ$ into the fourth quadrant.
\end{enumerate}

These effects can be real spatial anomalies (which would for point 2
be highly unlikely) or the result of the conversion itself under the,
at least in part, wrong assumptions of circular symmetry.

In particular the effect mentioned in point 1 could be reproduced if
the kinematic situation would be such that gas on a certain radius
from the Galactic center would give rise to a higher velocity along the line of sight in
the south than in the north, be this due to its own velocity or due
to that of the LSR\footnote{In the Hartmann \& Burton Leiden/Dwingeloo Survey the LSR was
defined in terms of the Standard Solar Motion of 20 km/s toward
$(\alpha,\delta)_{1900}=(18^{h},30^\circ)$.}.

Furthermore, the effect mentioned in point 2, when centered
symmetrically on $l=0^\circ$, is caused by local velocity-dispersed
gas, entering at the center direction into negative velocities for the
first quadrant and into positive velocities for the fourth and hence
accounted for originating from beyond the solar circle near
$l=0^\circ$ and scaled to densities accordingly (see also the comments
in the caption of Figure 13). This effect can also be seen in the
simulation of section 5.2. The displacement can be easily conceived of
by assuming a line of sight net positive velocity between the LSR and the
local gas in the center direction, be this due to local streaming or to a
radial velocity gradient by a non-circular global streaming.

To quantify all these conditions we measure parameters as follows. For
the items 4, 5, 6, and 7, we make use of the IDL routine GAUSSFIT as
described in section 4.1 to fit a Gaussian to the velocity
distribution in the $(l,v)$ plot at certain longitudes around the
cardinal directions $l=0^\circ$ respectively $l=180^\circ$, after we have
removed spectral features that are less than a factor e$^{-2}$
times the maximum $T_b$ value at that specific longitude, that
otherwise might have contaminated the outcome. The tightest velocity
constraints are then taken to be the minima along a plot of the
Gaussian widths varying with longitude, while the centroid of the
velocity distribution simply is the location of the center of the
Gaussian with longitude. For the items 1, 2, and 3, we only make use of
the contour in the $(l,v)$ display at $T_b=2$ K. Item 1 is the
maximum velocity along this contour over a broad longitude range
around $l=270^\circ$, respectively the minimum around
$l=90^\circ$. The longitudes where these extremes were recorded
represent item 2. A parameter, $v_{diff}$, is calculated from the
difference between the magnitude of the negative velocities at $l$,
for $0^\circ<l<180^\circ$, and the positive velocities at their
co--longitudes, at $360^\circ-l$. This parameter corresponds to the
shape of the extreme velocity curves, mentioned in item 3, and is
positive for larger velocities in the south than in the north at any
particular longitude. It is the same parameter that was used by Blitz
\& Spergel (1991)~\cite{bl:ell} as the principal indication of their
suggested outwardly decreasing ellipticity model. The emission mentioned
in item 8 is probably due to local gas that, in a purely circularly
symmetrical case, by its velocity dispersion would enter the (for its
circular kinematics) forbidden quadrants in equal amounts (as was
shown by the simulations of section five), but now has an extra
positive line of sight velocity. If the emission were due to gas from
larger radii in the second quadrant, shifted sufficiently to positive
velocities to show up at $v>0$ km/s, it might surprise why this
large--scale feature does not extend into the 1$^{st}$ and the
3$^{rd}$ quadrants, but instead disappears at their boundary quite
abruptly. Also, it is shown below in section seven that a net radial
inward (to the Galactic center) velocity of the LSR would reproduce
the desired effect, but this is strongly ruled out by other (global)
kinematic measures. Furthermore, a tangential error in the LSR would
either increase the extent to which the emission enters the formally
forbidden velocities for both quadrants or decrease them at the same
time. Finally, item 9 is quantified by integrating over velocities
with $v<0$ km/s for $0^\circ<l<180^\circ$ and with $v>0$ km/s for
their co--longitudes, at $360^\circ-l$. The $l$ values are then
subtracted from the ($360^\circ-l$) values and this is divided by the
($360^\circ-l$) values and then multiplied by 100 to give change in
column depth in percentages. The integration cutoff at $v=0$ km/s does
not significantly affect this parameter, $N_{diff}$, but instead the plot
remains quite similar if we take the cutoff at $v=\pm 10$, $\pm 20$ or
$\pm 30$ km/s, as was also shown by Blitz \& Spergel.

\begin{figure} 

\caption{Compilation of parameter plots derived from an $(l,v)$ display at $b=0^\circ$. The upper--left and middle plots show the net radial velocity of the peak 21 cm emission with respect to the LSR towards the center respectively anticenter. The lower--left and middle plots show the width ($\sigma$) of the total spectral distribution along any galactic longitude in the center respectively anticenter direction. The upper--right plot shows the difference in velocity of a 2 K contour of longitudes versus their co--longitudes ($360^\circ-l$), such that it is positive for larger velocities in the southern hemisphere. Finally, the lower--right plot is a measure of the percentage of excess column depth of the 4$^{th}$ quadrant with respect to the 1$^{st}$, and of the 3$^{rd}$ with respect to the 2$^{nd}$. In addition to this it followed that the most extreme velocity in the first two quadrants along a contour of 2 K in brightness temperature was $-126$ km/s occurring at $l=117^\circ$ ($27^\circ$ displaced from $l=90^\circ$), while for the last two it was $+144$ km/s occurring at $l=281^\circ$ ($11^\circ$ displaced from $l=270^\circ$), which renders their absolute ratio to be 1.1. Furthermore, emission from local gas enters along the same contour the formally forbidden second quadrant at an average velocity of 16.5 km/s, while in the third at $-11.0$ km/s, establishing an absolute ratio of 1.5.}
\end{figure}

\begin{figure} 

\caption{Same plots from an $(l,v)$--display at $|b|\leq 10^\circ$ as in Figure 24 are shown here. Again the upper--left and middle plots show the net radial velocity of the peak 21 cm emission with respect to the LSR towards the center respectively anticenter. The lower--left and middle plots show the width ($\sigma$) of the total spectral distribution along any galactic longitude in the center respectively anticenter direction. The upper--right plot shows the difference in velocity of a 2 K contour of longitudes versus their co--longitudes ($360^\circ-l$), such that it is positive for larger velocities in the southern hemisphere. Finally, the lower--right plot is a measure of the percentage of excess column depth of the 4$^{th}$ quadrant with respect to the 1$^{st}$, and of the 3$^{rd}$ with respect to the 2$^{nd}$. Corresponding values for the extreme velocity cutoffs at $T_b$=2 K are $-123$ km/s at $l=105^\circ$ ($15^\circ$ displaced from $l=90^\circ$) and 132 km/s at $l=277^\circ$ ($7^\circ$ displaced from $l=270^\circ$), with a ratio of 1.1. The local gas has averages of 15.2 km/s respectively $-12.2$, giving 1.2 as an absolute ratio.}
\end{figure}

Figure 24 shows a compilation of parameter plots derived from an
$(l,v)$ display at $b=0^\circ$ of the composite data cube and Figure
25 shows the same set of parameters for a latitude integrated $(l,v)$
data set in the range $|b|\leq 10^\circ$. Since the data grid points
are separated by half a degree in latitude, this last measurement
includes 41 cuts through the data of the sort that was used in the
first. The change in the plots from $b=0^\circ$ to $|b|\leq 10^\circ$
was monitored for the integrations $|b|\leq 1^\circ$, $|b|\leq
2^\circ$, and $|b|\leq 5^\circ$ as well and found to behave
continuously and to settle fast towards the situation of the $|b|\leq
10^\circ$ plots. It can thus be relied upon that this plot reflects
the global kinematic (and spatial) situation more accurately than any
of the others, moreover so because effects due to the deviation of the
HI layer from the plane $b=0^\circ$ at larger radii, are being
prevented, since emission at intensities large enough to have an
influence on the parameters under consideration here predominantly originates
from within the covered latitude range. The difference between the
plots of the centroid around $l=0^\circ$ can therefore be attributed to
a local irregularity in global streaming in the $b=0^\circ$ case,
something that is, in center direction, not implausible. It is reassuring for this reasoning that in the direction
$l=180^\circ$, where great kinematic turmoil is not observed, the
velocity centroid of both distributions have almost identical
positions. The greater spectral width of the plot of tightest
$v$--constraint with longitude around $l=0^\circ$ in the case of the
$b=0^\circ$ data cut with respect to the $|b|\leq 10^\circ$ version,
while in the direction of longitudes around $l=180^\circ$ they keep
identical values, is a feature that appears to be regular and inherent
to the dynamical situation and is reproduced in all of the simulations
we made in this project. In particular, note the difference between
Figure 27 and Figure 28, in this context. The change of the parameter
$v_{diff}$ along $0^\circ<|l|<180^\circ$ indicates that the
outer--Galaxy line of sight radial velocities, measured at T$_b$=2 K,
are more extreme in the 4$^{th}$ quadrant, and well into the 3$^{rd}$,
with respect to those from the 1$^{st}$ quadrant, and into the
2$^{nd}$. At certain directions ($l\approx110^\circ$ and
$l\approx250^\circ$) this reverses and the velocities in the 2$^{nd}$
quadrant become more extreme than those in the 3$^{rd}$. The column depth ($N_{diff}$) varies similarly. Here, around $l=100^\circ$,
there is a high relative peak to be seen of velocity integrated
densities in the southern data with respect to the northern, a
structure that has attributed to the Carina spiral arm by some
authors, as was mentioned by Blitz \& Spergel. In any case, this sort
of large structures seem to suggest that we should be cautious, when
we try to explain the shapes of these parameters by global kinematic
measures only, omitting other large scale influences like e.g. spiral
structure.

\begin{figure} 

\caption{Same plots from an $(l,v)$ display at $b=0^\circ$ as in Figure 24 are shown here for the simulated data according to a circularly symmetric model. Again the upper--left and middle plots show the net radial velocity of the peak 21 cm emission with respect to the LSR towards the center respectively anticenter. The lower--left and middle plots show the width ($\sigma$) of the total spectral distribution along any galactic longitude in the center respectively anticenter direction. The upper--right plot shows the difference in velocity of a 2 K contour of longitudes versus their co--longitudes ($360^\circ-l$), such that it is positive for larger velocities in the southern hemisphere. Finally, the lower--right plot is a measure of the percentage of excess column depth of the 4$^{th}$ quadrant with respect to the 1$^{st}$, and of the 3$^{rd}$ with respect to the 2$^{nd}$. Corresponding values for the extreme velocity cutoffs at $T_b$=2 K are $-156$ km/s at $l=90^\circ$ ($0^\circ$ displaced from $l=90^\circ$) and 156 km/s at $l=270^\circ$ ($0^\circ$ displaced from $l=270^\circ$), with a ratio of 1.0. The local gas has averages of 14.1 km/s respectively $-14.1$, giving 1.0 as an absolute ratio.}
\end{figure}

\begin{figure} 

\caption{Same plots from an $(l,v)$ display at $b=0^\circ$ as in Figure 24 are shown here for the simulated data according to a circularly kinematically symmetric model with a linear sinusoidal warp, with its maximum at $\theta=95^\circ$. Again the upper--left and middle plots show the net radial velocity of the peak 21 cm emission with respect to the LSR towards the center respectively anticenter. The lower--left and middle plots show the width ($\sigma$) of the total spectral distribution along any galactic longitude in the center respectively anticenter direction. The upper--right plot shows the difference in velocity of a 2 K contour of longitudes versus their co--longitudes ($360^\circ-l$), such that it is positive for larger velocities in the southern hemisphere. Finally, the lower--right plot is a measure of the percentage of excess column depth of the 4$^{th}$ quadrant with respect to the 1$^{st}$, and of the 3$^{rd}$ with respect to the 2$^{nd}$. Corresponding values for the extreme velocity cutoffs at $T_b$=2 K are $-133$ km/s at $l=98^\circ$ ($8^\circ$ displaced from $l=90^\circ$) and 136 km/s at $l=259^\circ$ ($-11^\circ$ displaced from $l=270^\circ$), with a ratio of 1.0. The local gas has averages of 14.1 km/s respectively $-14.1$, giving 1.0 as an absolute ratio.}
\end{figure}

\begin{figure} 

\caption{Same plots from an $(l,v)$ display at $|b|\leq 10^\circ$ as in Figure 24 are shown here for the simulated data according to a circularly kinematically symmetric model with a linear sinusoidal warp, with its maximum at $\theta=95^\circ$. Again the upper--left and middle plots show the net radial velocity of the peak 21 cm emission with respect to the LSR towards the center respectively anticenter. The lower--left and middle plots show the width ($\sigma$) of the total spectral distribution along any galactic longitude in the center respectively anticenter direction. The upper--right plot shows the difference in velocity of a 2 K contour of longitudes versus their co--longitudes ($360^\circ-l$), such that it is positive for larger velocities in the southern hemisphere. Finally, the lower--right plot is a measure of the percentage of excess column depth of the 4$^{th}$ quadrant with respect to the 1$^{st}$, and of the 3$^{rd}$ with respect to the 2$^{nd}$. Corresponding values for the extreme velocity cutoffs at $T_b$=2 K are $-135$ km/s at $l=95^\circ$ ($5^\circ$ displaced from $l=90^\circ$) and 135 km/s at $l=265^\circ$ ($-5^\circ$ displaced from $l=270^\circ$), with a ratio of 1.0. The local gas has averages of 14.1 km/s respectively $-14.1$, giving 1.0 as an absolute ratio.}
\end{figure}

\subsection{Symmetrical reference data}

As a reference to what would be the observational situation in case of
perfect circular symmetry or circular kinematic symmetry, but with a
linear sinusoidal warp, and as yet another check on the behavior of
the synthetic spectra, we performed the same asymmetry analysis as
above on the controlled models that were described in section
five. When considering perfect circular symmetry, there are indeed no
deviations detected of the sorts cataloged in the previous
subsection. This can be verified in Figure 26. This situation changes
slightly though, with respect to the parameters $v_{diff}$ and $N_{diff}$,
for the model including a linear sinusoidal warp (Figure 27). When we
integrate the $(l,v)$ display for $|b|\leq10^\circ$, however, this
difference almost totally vanishes, since then the warped layer is
included as well (Figure 28). Note, as was mentioned above, how this
also brings the two different tightest velocity constraints closer
together.

\subsection{Modeled asymmetries}

In general we can distinguish some categories of asymmetries with
respect to deviations from simple planar circularly symmetric rotation,
with the assumed flat rotation curve as described above.

\begin{itemize}
\item Deviations from planar rotation.
\item Deviations from a flat rotation curve.
\item Deviations in the LSR.
\item Deviations from circular streaming.  
\end{itemize}

The only deviation with immediate observational consequences that is
quite firmly established is the one that is first mentioned and has
for a symmetrical warp been tested already in the previous
subsection. The possibilities that were further investigated here were the
following:
\begin{itemize}
\item Asymmetrical warp
\item Radial velocity LSR inward/outward from the GC.
\item Global circular streaming, with sun on elliptical movement.
\item Global elliptical streaming.
\item Small radial velocity of local gas in the solar neighborhood.
\item Linear change in ellipticity from solar radius outward.
\end{itemize}

With respect to deviations from a flat rotation curve, we can comment as
follows. A linear or monotonically and smoothly varying rotation
curve will cause only scaling effects due to the different
velocity--to--distance transformation. Local perturbations with $R$,
however, or, for that matter, with $\theta$, certainly will cause the
appearance of the spectra and, after conversion, that of the derived
Galactic morphology, to change significantly. Since we consider global
parameters only here, we did not perform such simulations. During our
investigations various different values for the parameters that occur
in the models were tried out to monitor the response of the generated
spectra to these changes, as will be described in the next section.

\section{Observational implications of model asymmetries}

After having cataloged, in the previous section, a set of observed
asymmetries and constraints and listed a number of morphologies and
kinematics that could mimic these asymmetries, we will proceed to
pursue the observational implications of certain specific models for
the observed spectra. It has been shown which properties of the
spectra require that the assumptions of circular spatial and
kinematic symmetry be abandoned. In this section it will become apparent which
modeled asymmetries produce spectra that point in the right
direction. For clearness, we will monitor one change with respect to
the standard model of section 5.1 at a time, except in 7.7, when we
consider global elliptical streaming in combination with an
asymmetrical warp. This, to give galactocentric cylindrical plots that
are comparable with the real data, since the asymmetrical warp is
quite firmly observationally established (and it is the only one in
this section to have been so).  We are putting limits on a possible
error in the determination of the motion of the Local Standard of Rest
as well as on possible ellipticity in combination with different
orientations of the semi--major axis with respect to the line
connecting the sun with the Galactic center. Such galactic asymmetries
could mimic some of the observed bilateral asymmetries in the
$(l,b,v)$ data. We conclude this section with a consideration of the
effects of a velocity--to--distance conversion, using circular
kinematics, on the ($l,b,v$) spectra of a Galaxy that, in fact,
rotates along global elliptical streamlines.

\subsection{Circular symmetry with asymmetrical warp}

The first deviation from the model in section 5.1 is that of an
asymmetrical warp. It should be noted that this deviation is less
hypothetical than the others that will follow, since it has been
accurately measured. Even though this measurement was performed in the
data cube that was constructed using circular kinematics instead of
some possibly better, but yet unknown, alternative, it cannot be
expected to differ more than of the order of, at most, ten percent
from reality. The shape of the warp is again sinusoidal, but this time
the amplitude differs in the part of the hemisphere that is accessible
from the north from that of the south, according to the curves that
are plotted in Figure 16. The maximum of displacement from the plane
$b=0^\circ$ was set to be attained at $\theta=120^\circ$. It was
separated from the minimum by $180^\circ$.

An ($l,v$) plot of the resulting spectra is shown in Figure 29. It is
clearly seen that in the 3$^{rd}$ and 4$^{th}$ quadrants, by the
folding--back property of the asymmetrical warp in the south,
velocities are recorded along the lowest brightness temperature
contours that are more extreme than those of the 1$^{st}$ and 2$^{nd}$
quadrants. Furthermore, in the compilation of parameter plots, that
check for asymmetry, in Figure 30, it is interesting to note the large
impact this has on the parameters $v_{diff}$ and $N_{diff}$, and on the
ratio of the extreme velocity cutoffs, for a cross--cut at
$b=0^\circ$. Taking into account the relative certainty with which we
should accept the asymmetrical warp effects, they should be imagined
to reside implicitly in the same real data's $b=0^\circ$ parameter
plots. However, it was demonstrated earlier (in section 6.2) that for a
$|b|\leq 10^\circ$ plot these effects largely disappear. Also, mind
that the distortions of the above mentioned parameters from symmetry
due to the asymmetrical warp will have to be superimposed on those
resulting from the following investigations, if comparison is to
be made with the plots in Figure 24. Particularly, it is encouraging
that the $N_{diff}$ parameter in this figure is quite indented
for the first few tens of degrees in $|l|$, when compared with its
$|b|\leq 10^\circ$ counterpart, consistent with the curve below 0\%
of $N_{diff}$ in Figure 30. To visualize its asymmetrical character,
the warp was also displayed in a cylindrical cut through $z$ and
$\theta$ at the galactocentric distance of $R=20$ kpc.

\begin{figure} 

\caption{Composite spectral image at $b=0^\circ$, for a model with circular kinematical symmetry, but asymmetrical warp with its maximum at $\theta=120^\circ$, as described in the text. As in Figure 1, the peak intensity of the upper panel display corresponds to a brightness temperature of 135 K and the lightest grey--scale corresponds to $T_b$=4 K.}
\end{figure}

\begin{figure} 

\caption{Same plots for $b=0^\circ$ as in Figure 24, for the case of circular kinematical symmetry with an asymmetrical warp with a maximum at $\theta=120^\circ$. Again the upper--left and middle plots show the net radial velocity of the peak 21 cm emission with respect to the LSR towards the center respectively anticenter. The lower--left and middle plots show the width ($\sigma$) of the total spectral distribution along any galactic longitude in the center respectively anticenter direction. The upper--right plot shows the difference in velocity of a 2 K contour of longitudes versus their co--longitudes ($360^\circ-l$), such that it is positive for larger velocities in the southern hemisphere. Finally, the lower--right plot is a measure of the percentage of excess column depth of the 4$^{th}$ quadrant with respect to the 1$^{st}$, and of the 3$^{rd}$ with respect to the 2$^{nd}$. Corresponding values for the extreme velocity cutoffs at $T_b$=2 K are $-111$ km/s at $l=89^\circ$ ($-1^\circ$ displaced from $l=90^\circ$) and 156 km/s at $l=270^\circ$ ($0^\circ$ displaced from $l=270^\circ$), with a ratio of 1.4. The local gas has averages of 14.1 km/s respectively $-14.1$, giving 1.0 as an absolute ratio.}
\end{figure}

\begin{figure} 

\caption{Mean HI volume densities in a cylindrical cut through the simulated 
data cube at a galactocentric distance of $R=20$ kpc, for a model with circular kinematical symmetry, but asymmetrical warp as described in the text. The warp maximum here was set at $\theta=120^\circ$. From the intersection of the Galactic midplane with the plane established by $b=0^\circ$ it can be seen that the line of nodes resulting from this warp maximum in the case of a sinusoidal warp is not consistent with what is seen in the real data (Figures 2--5).}
\end{figure}

\subsection{Radial velocity LSR}

Superimposed upon the standard flat plane--parallel simulated Galaxy
with circular kinematic and spatial symmetry (section 5.1) is the
effect of a 10 km/s motion of the LSR radially outward from the
Galactic center. Figure 32 shows the spectral situation of this model,
again for $b=0^\circ$. From this figure and from the compilation of
parameter plots in Figure 33 it can clearly be seen that both in the
center and in the anticenter directions the centroid of the spectrum has an
off--set exactly as great as the initial radial velocity in absolute
magnitude. The $v_{diff}$ parameter shows a reasonable fit to the
real data (Figure 25), although a slightly better fit would be
attained for a higher radial outward velocity. This is exactly what
Blitz \& Spergel derived for the same parameter in their
simulations. As a best fit they found $v_{rad}$=14 km/s
outward. However, our other parameters strongly oppose such an
outcome. For, not only would this increase the velocity of the
centroid with respect to the center direction, which is not desirable
in view of Figure 25, but also, and this is most important, does it
contradict the observational evidence of little or no velocity with
respect to the anticenter of the spectral centroid found there. On the
other hand it is possible to reconcile $N_{diff}$ with the
observations, but it will be shown below that this is also possible
for other models. The same is true for $v_{diff}$. Finally, by itself not conclusive, but still a matter of concern, is that the ratio
of the average local gas velocities shows a directly opposite movement
to what is observed. Note, that in case of an inward velocity all
effects would be the same, but reversed. Through the above mentioned
reasoning we rule out any correction to the LSR used in the
Leiden/Dwingeloo survey greater than a few km/s.

\begin{figure} 

\caption{Composite spectral image at $b=0^\circ$, for a model with circular symmetry but a radial outward velocity of 10 km/s as described in the text. As in Figure 1, the peak intensity corresponds to a brightness temperature of 135 K and the lightest grey--scale corresponds to $T_b$=4 K.}
\end{figure}

\begin{figure} 

\caption{Same plots for $b=0^\circ$ as in Figure 24, for the case of circular symmetry with a radial outward velocity of 10 km/s. Again the upper--left and middle plots show the net radial velocity of the peak 21 cm emission with respect to the LSR towards the center respectively anticenter. The lower--left and middle plots show the width ($\sigma$) of the total spectral distribution along any galactic longitude in the center respectively anticenter direction. The upper--right plot shows the difference in velocity of a 2 K contour of longitudes versus their co--longitudes ($360^\circ-l$), such that it is positive for larger velocities in the southern hemisphere. Finally, the lower--right plot is a measure of the percentage of excess column depth of the 4$^{th}$ quadrant with respect to the 1$^{st}$, and of the 3$^{rd}$ with respect to the 2$^{nd}$. Corresponding values for the extreme velocity cutoffs at $T_b$=2 K are $-157$ km/s at $l=93^\circ$ ($3^\circ$ displaced from $l=90^\circ$) and 157 km/s at $l=273^\circ$ ($3^\circ$ displaced from $l=270^\circ$), with a ratio of 1.0. The local gas has averages of 7.4 km/s respectively $-20.9$, giving 0.4 as an absolute ratio.}
\end{figure}

\subsection{Solar elliptical movement}

\begin{figure} 

\caption{Construction of the rotational situation used in the solar elliptical movement simulation. For clarity, the ellipticity is greatly exaggerated.}
\end{figure}

In the case of a solar elliptical movement, but keeping the remainder
of the Galaxy identical to that in section 5.1, the effects mimic in
part those for a solar radial velocity. The rotational situation is
shown in Figure 34 for a face--on view of the Galaxy as seen from the
north pole. In Figures 35 and 36 the results are shown for the case
of an ellipticity of $\epsilon$=0.05 and an angle between the
sun--center line and the semi--major axis of its orbit of
$\alpha=45^\circ$. The line of sight component of the sun's orbital
velocity immediately creates a non--acceptable shift with respect to
the anticenter centroid and even a double shift in the center
direction. At the same time the ratio of the local gas average
velocities again moves in the wrong direction. Any measure to reduce
both shifts in order to get within reach of the real observables,
reduces the effects on $v_{diff}$ and $N_{diff}$ correspondingly and
thus seems not very fruitful. These measures include a decrease in
ellipticity and a change of $\alpha$ towards both $0^\circ$ and
$90^\circ$. For these reasons, we rule out the solar elliptical
movement according to the model as discussed as a natural explanation
of the observed asymmetries.

We must note that all alternatives that imply a motion of the LSR with
respect to its immediate neighborhood are likely to be refuted by the
same constraints mentioned above, since the bigger part of all the 21 cm intensity is emitted by gas that resides within a few kpc distance from the sun. The parameter
$v_{diff}$ on the other hand originates from the outer Galaxy only. As
we shall see, is in the case of global elliptical streaming the
velocity difference with respect to the immediate solar neighborhood
limited, whereas with respect to the region outside the sun--center
radius in the direction of $l=0^\circ$ it is considerable. This is in
direct correspondence with the observed data.

\begin{figure} 

\caption{Composite spectral image at $b=0^\circ$, for a model with circular kinematics, but with the sun moving on a slightly elliptical orbit with $\epsilon=0.05$ and $\alpha=45^\circ$, as described in the text. As in Figure 1, the peak intensity of the upper panel display corresponds to a brightness temperature of 135 K and the lightest grey--scale corresponds to $T_b$=4 K.}
\end{figure}

\begin{figure} 

\caption{Same plots for $b=0^\circ$ as in Figure 24, for the case of circular kinematics with the sun moving on a slightly elliptical orbit, with $\epsilon=0.05$ and $\alpha=45^\circ$. Again the upper--left and middle plots show the net radial velocity of the peak 21 cm emission with respect to the LSR towards the center respectively anticenter. The lower--left and middle plots show the width ($\sigma$) of the total spectral distribution along any galactic longitude in the center respectively anticenter direction. The upper--right plot shows the difference in velocity of a 2 K contour of longitudes versus their co--longitudes ($360^\circ-l$), such that it is positive for larger velocities in the southern hemisphere. Finally, the lower--right plot is a measure of the percentage of excess column depth of the 4$^{th}$ quadrant with respect to the 1$^{st}$, and of the 3$^{rd}$ with respect to the 2$^{nd}$. Corresponding values for the extreme velocity cutoffs at $T_b$=2 K are $-161$ km/s at $l=92^\circ$ ($2^\circ$ displaced from $l=90^\circ$) and 151 km/s at $l=272^\circ$ ($2^\circ$ displaced from $l=270^\circ$), with a ratio of 0.9. The local gas has averages of 10.5 km/s respectively $-17.6$, giving 0.6 as an absolute ratio.}
\end{figure}

\subsection{Global elliptical streaming}

\begin{figure} 

\caption{Construction of the rotational situation used in the global elliptical streaming simulation. For clearness, the ellipticity is greatly exaggerated.}
\end{figure}

The model incorporating global elliptical streaming is further
clarified in Figure 37. Reviewed here is the situation for
$\epsilon$=0.05 and $\alpha=45^\circ$. Figures 38 and 39 show the
results of its analysis. It is immediately seen that this model
combines positive results for all parameters. Thus, it causes the
required positive velocity displacement for the centroid in the center
direction, and still keeps that in the anticenter direction fixed at
zero. Both $v_{diff}$ and $N_{diff}$ show a distortion that can, in
combination with that of the asymmetrical warp, be united with the
situation for the real data. On top, the ratio of the local gas
average velocities agrees more here to that observed. We cannot
give a fixed set of variables here that represent a best fit, but it
is clear that this is the only one of our models that accounts for the general
trends without contradicting others. In fact, it has not been found
possible by any other model than global elliptical streaming, through
the simulations we used, to reproduce the required positive velocity
of the center direction centroid and at the same time keep the
centroid in the anticenter direction fixed at $v$=0 km/s within a few
km/s. Finally, if the angle $\alpha$ were to be set at negative values
all plots would show the same curves, but inverted. It is interesting
to see that the conclusion resulting from this subsection invokes a
ellipticity and an orientation of the inflicted morphology, that is
consistent with the model that Blitz \& Spergel put forward. Although
it can be seen that their assumption, that the kinematic influence on
$v_{diff}$ of a radial component of the solar motion along elliptical
streamlines is identical to a true radial velocity of the LSR, is not
generally valid.

\begin{figure} 

\caption{Composite spectral image at $b=0^\circ$, for a model with global elliptical streaming as described in the text. As in Figure 1, the peak intensity of the upper panel display corresponds to a brightness temperature of 135 K and the lightest grey--scale corresponds to $T_b$=4 K.}
\end{figure}

\begin{figure} 

\caption{Same plots for $b=0^\circ$ as in Figure 24, for the case of global elliptical streaming. Again the upper--left and middle plots show the net radial velocity of the peak 21 cm emission with respect to the LSR towards the center respectively anticenter. The lower--left and middle plots show the width ($\sigma$) of the total spectral distribution along any galactic longitude in the center respectively anticenter direction. The upper--right plot shows the difference in velocity of a 2 K contour of longitudes versus their co--longitudes ($360^\circ-l$), such that it is positive for larger velocities in the southern hemisphere. Finally, the lower--right plot is a measure of the percentage of excess column depth of the 4$^{th}$ quadrant with respect to the 1$^{st}$, and of the 3$^{rd}$ with respect to the 2$^{nd}$. Corresponding values for the extreme velocity cutoffs at $T_b$=2 K are $-166$ km/s at $l=90^\circ$ ($0^\circ$ displaced from $l=90^\circ$) and 147 km/s at $l=271^\circ$ ($1^\circ$ displaced from $l=270^\circ$), with a ratio of 0.9. The local gas has averages of 14.1 km/s respectively $-14.1$, giving 1.0 as an absolute ratio.}
\end{figure}

\subsection{Local streaming}

Since there was no global model found that could naturally explain the
greater than one value of the absolute ratio of local gas average
velocities without going against other strong constraints, a non--global
explanation was thought of that could. This was that of local
streaming. The model is again the Galaxy of section 5.1, only this
time with an area around the sun's location within a radius of 250 pc
that has a velocity radially outward from the Galactic center of 7
km/s. It is important to realize that by our parameters in the
compilation of Figure 41 only the response of the data in the anticenter
direction is monitored, and that, in fact, in the center direction the
reasoning around the asymmetrical, elongated patch of gas (section
6.1) would argue for a motion away from the sun. In Figures 40 and 41
it can be seen that local streaming reaches the right objectives and
yet leaves the other observables almost unaffected. When we increase
the radius of the local area that is given a net velocity, the parameters
increasingly respond to it in accordance with an error in the LSR, as
more overall HI intensity is taken to have a distorted velocity. As a local effect one would need to incorporate all local effects that are associated with
the solar neighborhood on such scales, in order to derive valid
conclusions about Galactic properties. As such it is not meant to describe the solar neighborhood in this section. However this model is taken to
be suggestive of a remarkability in the analysis of the parameters we monitor, in view of an observed asymmetry.

\begin{figure} 

\caption{Composite spectral image at $b=0^\circ$, for a model with local streaming as described in the text. As in Figure 1, the peak intensity corresponds to a brightness temperature of 135 K and the lightest grey--scale corresponds to $T_b$=4 K.}
\end{figure}

\begin{figure} 

\caption{Same plots for $b=0^\circ$ as in Figure 24, for the case of local streaming. Again the upper--left and middle plots show the net radial velocity of the peak 21 cm emission with respect to the LSR towards the center respectively anticenter. The lower--left and middle plots show the width ($\sigma$) of the total spectral distribution along any galactic longitude in the center respectively anticenter direction. The upper--right plot shows the difference in velocity of a 2 K contour of longitudes versus their co--longitudes ($360^\circ-l$), such that it is positive for larger velocities in the southern hemisphere. Finally, the lower--right plot is a measure of the percentage of excess column depth of the 4$^{th}$ quadrant with respect to the 1$^{st}$, and of the 3$^{rd}$ with respect to the 2$^{nd}$. Corresponding values for the extreme velocity cutoffs at $T_b$=2 K are $-156$ km/s at $l=90^\circ$ ($0^\circ$ displaced from $l=90^\circ$) and 156 km/s at $l=270^\circ$ ($0^\circ$ displaced from $l=270^\circ$), with a ratio of 1.0. The local gas has averages of 18.2 km/s respectively $-11.0$, giving 1.6 as an absolute ratio.}
\end{figure}

\subsection{Decreasing versus constant ellipticity}

Finally, we comment upon the difference of constant ellipticity versus
decreasing ellipticity, as advocated by Blitz \& Spergel. The same
global elliptical model as in the subsection 7.5 above was altered to
let the ellipticity decrease linearly beyond $R=$9 kpc to zero at
$R=$25 kpc. A parameter compilation plot was made, as shown in Figure
42, comparable to that in Figure 39. The slight differences might be
pleading in favor of the decreasing ellipticity, though not very
strongly on the basis of our analysis. For comparison, we tried a
version with increasing ellipticity from the solar radius onwards,
with negative results.

\begin{figure} 

\caption{Same plots for $b=0^\circ$ as in Figure 24, for the case of an outward decreasing elliptical streaming. Again the upper--left and middle plots show the net radial velocity of the peak 21 cm emission with respect to the LSR towards the center respectively anticenter. The lower--left and middle plots show the width ($\sigma$) of the total spectral distribution along any galactic longitude in the center respectively anticenter direction. The upper--right plot shows the difference in velocity of a 2 K contour of longitudes versus their co--longitudes ($360^\circ-l$), such that it is positive for larger velocities in the southern hemisphere. Finally, the lower--right plot is a measure of the percentage of excess column depth of the 4$^{th}$ quadrant with respect to the 1$^{st}$, and of the 3$^{rd}$ with respect to the 2$^{nd}$. Corresponding values for the extreme velocity cutoffs at $T_b$=2 K are $-160$ km/s at $l=92^\circ$ ($2^\circ$ displaced from $l=90^\circ$) and 160 km/s at $l=272^\circ$ ($2^\circ$ displaced from $l=270^\circ$), with a ratio of 1.00. The local gas has averages of 14.0 km/s respectively $-14.2$, giving 1.0 as an absolute ratio.}
\end{figure}

\subsection{Conversion remarkabilities}

There is a compelling argument that could make it worthwhile to test a
conversion to galactocentric ($R,\theta,z$) coordinates from
heliocentric ($l,b,v$) coordinates taking as an input model a
combination of the best options of the alternatives that we went
through in the subsections above. This would be a combination of the
asymmetrical warp and the global elliptical streaming with
$\alpha=45^\circ$ and an ellipticity of $\epsilon$=0.05. This argument
is, of course, that if our Galaxy were to be of such morphology, with
corresponding kinematics, then at the velocity--to--distance
conversion, assuming the wrong type of kinematics, errors would be
made with respect to the distances that are then attributed to certain
emissivities and, what is at least as bad, volume densities are calculated accordingly. As in section 5.2, we thus constructed the
complete ($l,b,v$) data cube for the combined model mentioned
here. Resulting plots for spectra through $v$ and $l$ for $b=0^\circ$
and for $|b|\leq 10^\circ$ are shown in Figure 43. Compare the two
compilation parameter plots in Figures 44 and 45 that were derived
from them. It is instructive to note that the velocity of the
centroid of the spectrum in the center direction is lessened significantly
when averaged over several (41 to be exact) cross--cuts in
latitude. Compared to Figure 25, this ellipticity suddenly does not
seem very large anymore, but instead quite appropriate. Also, note,
how the effect of the asymmetrical warp is greatly reduced for the
case $|b|\leq 10^\circ$. Now let us consider the remarkabilities
concerning the velocity--to--distance data processing. As one can see
in Figure 46, the discrepancy between the input model and its display
in ($R,\theta,z$) after conversion is by far greater than that, as
described in section 5.2.2, of the case of identical model and
conversion kinematics. Since, in addition to some of the vagaries that
were commented upon there, we see some striking new ones. First, it appears that,
just as in the Figure 3--6 displays, the 4$^{th}$ quadrant is
overabundant in gas with respect to the 1$^{st}$ and similarly
the 2$^{nd}$ quadrant displays a higher volume density than the
3$^{rd}$, especially for larger radii. Secondly, the rise to high $z$
from the 2$^{nd}$ quadrant to the 1$^{st}$ of the HI gas layer and its
subsequent detachment of that in the 4$^{th}$ quadrant. A feature
that, again more towards higher radii, is also seen in the real
data. Finally, the apparent larger overall volume densities of the
southern hemisphere data with respect to the northern. This is
perceived to be due to the kinematic effect that was explained in section
6.1.

\begin{figure} 

\vfill 
%\includegraphics[bb=65 365 550 720, width=13cm, height=9cm]{figure43b.ps}
%\psfig{file=/strw11/voskes/diagrams/ellaswarp/lvbintabs10m95.ps,width=14cm,height=9cm} 
%\vskip 17cm
\caption{(l,v) plot at $b=0^\circ$, for the simulated data with elliptical streaming and asymmetrical warp with its maximum at $\theta=95^\circ$ (upper panel), as well as a latitude integrated version $|b|\leq 10^\circ$ (lower panel), to show the effects this has on the influence of the warp. As in Figure 1, the peak intensity of the upper panel display corresponds to a brightness temperature of 135 K and the lightest grey--scale corresponds to $T_b$=4 K. The lower panel display was averaged over its $b$ values and has a deviating grey--scale scaled between minimum and maximum intensities.}
\end{figure}

\begin{figure} 

\caption{Same plots from an $(l,v)$ display at $b=0^\circ$ as in Figure 24 are shown here for the simulated data according to a model with elliptical streaming and asymmetrical warp with its maximum at $\theta=95^\circ$. Again the upper--left and middle plots show the net radial velocity of the peak 21 cm emission with respect to the LSR towards the center respectively anticenter. The lower--left and middle plots show the width ($\sigma$) of the total spectral distribution along any galactic longitude in the center respectively anticenter direction. The upper--right plot shows the difference in velocity of a 2 K contour of longitudes versus their co--longitudes ($360^\circ-l$), such that it is positive for larger velocities in the southern hemisphere. Finally, the lower--right plot is a measure of the percentage of excess column depth of the 4$^{th}$ quadrant with respect to the 1$^{st}$, and of the 3$^{rd}$ with respect to the 2$^{nd}$. Corresponding values for the extreme velocity cutoffs at $T_b$=2 K are $-133$ km/s at $l=98^\circ$ ($8^\circ$ displaced from $l=90^\circ$) and 136 km/s at $l=259^\circ$ ($-11^\circ$ displaced from $l=270^\circ$), with a ratio of 1.0. The local gas has averages of 14.1 km/s respectively $-14.1$, giving 1.0 as an absolute ratio.}
\end{figure}

\begin{figure} 

\caption{Same plots from an $(l,v)$ display at $|b|\leq 10^\circ$ as in Figure 24 are shown here for the simulated data according to a model with elliptical streaming and asymmetrical warp with its maximum at $\theta=95^\circ$. Again the upper--left and middle plots show the net radial velocity of the peak 21 cm emission with respect to the LSR towards the center respectively anticenter. The lower--left and middle plots show the width ($\sigma$) of the total spectral distribution along any galactic longitude in the center respectively anticenter direction. The upper--right plot shows the difference in velocity of a 2 K contour of longitudes versus their co--longitudes ($360^\circ-l$), such that it is positive for larger velocities in the southern hemisphere. Finally, the lower--right plot is a measure of the percentage of excess column depth of the 4$^{th}$ quadrant with respect to the 1$^{st}$, and of the 3$^{rd}$ with respect to the 2$^{nd}$. Corresponding values for the extreme velocity cutoffs at $T_b$=2 K are $-135$ km/s at $l=95^\circ$ ($5^\circ$ displaced from $l=90^\circ$) and 135 km/s at $l=265^\circ$ ($-5^\circ$ displaced from $l=270^\circ$), with a ratio of 1.0. The local gas has averages of 14.1 km/s respectively $-14.1$, giving 1.0 as an absolute ratio.}
\end{figure}

\begin{figure} 

\vfill 

%\includegraphics[bb=65 365 550 720, width=13cm, height=8cm]{figure46b.ps}
%\psfig{file=/strw11/voskes/diagrams/ellaswarp/r26.ps,width=14cm,height=9cm} 
%\vskip 17cm
\caption{Mean HI volume densities in cylindrical cuts through the simulated 
data cube with elliptical streaming and asymmetrical warp with its maximum at $\theta=95^\circ$ at galactocentric distances of $R=16$ kpc (upper) and 26 kpc (lower).
The cylindrical--coordinate $(R,\theta,z)$ data set was sampled at $1^\circ$ 
intervals in $\theta$ and 25--pc intervals in $z$; the velocity--to--distance 
transformation displayed here was circularly symmetric. The darkest grey--scale was scaled to the maximum densities at each of both radii, to be able to show details even at $R$=26 kpc. For reference a line tracing the galactic plane at $b=0^\circ$ was drawn. The warp maximum here was set at $\theta=95^\circ$.}
\end{figure}

\section{Summary}

In this project the Leiden/Dwingeloo data was used as a main
constituent in a composite data cube to display the outer reaches of
our Galaxy. To account for any possible distortion of the warp
description due to anomalous velocity gas the principle HVC complexes
as identified by Wakker \& van Woerden were masked out before the
velocity--to--distance conversion took place. For this conversion we
used the IAU standards of R$_0$=8.5 kpc and $\Theta_0$=220
km/s. Due to the properties of the Leiden/Dwingeloo survey, at the
same time a more sensitive and more detailed description of the warp
and its flaring nature was given than had been possible before. A
range of warp parameters was further quantified and compared to
previous investigations. In order to follow the method of converting
from heliocentric ($l,b,v$) coordinates to galactocentric
($R,\theta,z$) coordinates a complete set of synthetic spectra was
constructed on the same grid as the Leiden/Dwingeloo survey and
subsequently processed in identical manner as the real data. Several
observational vagaries that resulted from this test conversion have
been identified in the real spectra. In addition controlled
conversion has been done for a range of hypothetical Galaxy models to
search for resulting remarkabilities that can be associated with the
observational situation, since it has been long established that the
spectra show asymmetries with respect to global circular spatial and
kinematic symmetry. In order to monitor this quantitatively, we choose
a fixed set of parameters in the ($l,b,v$) data and plotted their
measured values in a single compilation display. This was done for the
real data as a reference as well. It appeared that many different
models create an effective significant motion of the LSR, which is by
our parameter plots strongly confined to very low velocities. Judging
on a global scale these models can therefore be ruled out. To the
observational situation, both directly in heliocentric coordinates and
in a more subtle (kinematical) way in galactocentric coordinates
(after conversion using circular symmetry), remarks can be made that
agree with those that follow from a modeled elliptical input
Galaxy. This outer Galaxy ellipticity is consistent with that found by
Blitz \& Spergel. We note, furthermore, that it has the same
orientation as the gas streamlines that resulted from the 
modeling of the motions in the inner few kpc by Liszt \& Burton
(1980)~\cite{li:arm}


\begin{thebibliography}{99}

\bibitem{ba:lop} Baldwin, J.E., Lynden--Bell, D., Sancisi, R. 1980, {\em Monthly Notices Roy. Astron. Soc. } 193, 313
\bibitem{bl:ell}Blitz, L., Spergel, D.N. 1991, {\em Astrophys. J. } 370, 205
\bibitem{bk:warp}Burke, B.F. 1957, {\em Astron. J. } 62, 90
\bibitem{bu:pro}Burton, W.B. 1972, {\em Astron. Astrophys. } 19, 51--65
\bibitem{bu:lop}Burton, W.B. 1973, {\em Pub. Astron. Soc. Pacific } 85, 679
\bibitem{bu:HI}Burton, W.B. 1985, {\em Astron. Astrophys. Suppl. Ser. } 62, 365
\bibitem{bu:warp}Burton, W.B., te Lintel Hekkert, P. 1986, {\em Astron. Astrophys. Suppl. Ser. } 65, 427
\bibitem{bu:ell}Burton, W.B. 1991, {\em The Galactic Interstellar Medium, Saas--Fee Advanced Course 21, ed. D. Pfenniger and P. Bartholdi (Berlin: Springer--Verlag)}, 126
\bibitem{di:warp}Diplas, A., Savage, B.D. 1991, {\em Astrophys. J. } 377, 126
\bibitem{ha:gal}Hartmann, D., Burton, W.B. 1997, {\em Atlas of Galactic Neutral Hydrogen (Cambridge: Cambridge University Press)} 
\bibitem{he:galpar}Henderson, A.P., Jackson, P.D., Kerr, F.J. 1982, {\em Astrophys. J. } 263, 116--122
\bibitem{ke:warp}Kerr, F.J. 1957, {\em Astron. J. } 62, 93
\bibitem{ke:ell}Kerr, F.J. 1962, {\em Monthly Notices Roy. Astron. Soc. } 123, 327--45
\bibitem{ke:HI}Kerr, F.J., Bowers, P.F., Jackson, P.D., Kerr, M. 1986, {\em Astron. Astrophys. Suppl. Ser. } 66, 373
\bibitem{kt:ell}Kuijken, K., Tremaine, S. 1991, {\em Dynamics of Disc Galaxies (Sweden: Varberg Castle)}, 71
\bibitem{ku:ell}Kuijken, K. 1992, {\em Pub. Astron. Soc. Pacific} 104, 809--811
\bibitem{ku:scal}Kulkarni, S.R., Blitz, L., Heiles, C. 1982, {\em Astrophys. J. } 259, L63--L66
\bibitem{li:arm}Liszt, H.S., Burton, W.B. 1980, {\em Astrophys. J. } 236, 779
\bibitem{li:HI}Liszt, H.S., Burton, W.B. 1983, {\em Astron. Astrophys. Suppl. Ser. } 52, 63
\bibitem{oo:warp}Oort, J. H., Kerr, F.J., Westerhout, G. 1958, {\em Monthly Notices Roy. Astron. Soc. } 118, 379
\bibitem{ri:lop} Richter, O.-G., Sancisi, R. 1994, {\em Astron. Astrophys. } 290, L9
\bibitem{st:HI}Stark, A.A., Gammie, C.F., Wilson, R.W., Bally, J., Linke, R.A., Heiles, C., Hurwitz, M. 1992, {\em Astrophys. J. Suppl. Ser. } 79, 77
\bibitem{sw:lop} Swaters, R.A., Schoenmakers, R.H.M., Sancisi, R., van Albada, T.S. 1999, {\em Monthly Notices Roy. Astron. Soc. } 304, 330
\bibitem{wa:hvc}Wakker, B.P., van Woerden, H. 1997, {\em Annu. Rev. Astron. Astrophys. } 35, 217
\bibitem{wi:s7}Williams, D.R.W. 1973, {\em Astron. Astrophys. Suppl. } 8, 505
\bibitem{wo:galpar}Wouterloot, J.G.A., Brand, J., Burton, W.B., and Kwee, K.K. 1990, {\em Astron. Astrophys. } 230, 21--36

\end{thebibliography}
\end{document}